\documentclass[aps,pra, twocolumn,superscriptaddress,longbibliography]{revtex4-2}
\usepackage{graphicx}
\usepackage{float}

\usepackage{derivative}
\usepackage{xcolor}
\usepackage[margin=0.68in]{geometry}
\usepackage{siunitx}
\usepackage{soul}
\usepackage{minitoc}
\usepackage{color}
\usepackage{amssymb}
\usepackage{amsmath, scalerel, breqn}
\usepackage{bbm}
\usepackage{physics}
\usepackage{bbold}
\usepackage{enumitem}

\definecolor{dark-red}{rgb}{0.4,0.15,0.15}
\definecolor{dark-blue}{rgb}{0.15,0.15,0.4}
\definecolor{medium-blue}{rgb}{0,0,0.5}
\usepackage{hyperref}
\hypersetup{
    colorlinks, linkcolor={dark-red},
    citecolor={dark-blue}, urlcolor={medium-blue}
}

\definecolor{green}{rgb}{0.08,0.7,0.05}

\usepackage[normalem]{ulem}
\begin{document}
\doparttoc 
\faketableofcontents 
\part{} 

\setcitestyle{super}

\title{Measurement-induced entanglement and teleportation on a noisy quantum processor}

\author{Google Quantum AI and Collaborators$^{\hyperlink{authorlist}{\dagger},}$}
\date{\today}

\email[Corresponding authors: V.~Khemani: ]{vkhemani@stanford.edu}
\email[\\ P. Roushan: ]{pedramr@google.com}

\begin{abstract} 
Measurement has a special role in quantum theory~\cite{Solvay2011}: by collapsing the wavefunction it can enable phenomena such as teleportation~\cite{Bennett_teleportation_1993} and thereby alter the ``arrow of time'' that constrains unitary evolution. When integrated in many-body dynamics, measurements can lead to emergent patterns of quantum information in \emph{space-time}~\cite{SkinnerPRX2019, LiPRB2018, GullansPRX, ChoiPRL2020,JianPRB2020,Barkeshli_NatPhys_2021,Ippoliti2021postselection,FanPRB2021} that go beyond established paradigms for characterizing phases, either in or out of equilibrium~\cite{landau_ginzburg, wen_topologicalorder_1990, sondhi_floquetreview_2020}. 
On present-day NISQ processors~\cite{preskill2018quantum}, the experimental realization of this physics is challenging due to hardware limitations and the stochastic nature of quantum measurement. Here we address these experimental challenges and study measurement-induced quantum information phases on up to 70 superconducting qubits. By leveraging the interchangeability of space and time, we use a duality mapping~\cite{Ippoliti2021postselection, Lu2021spacetime, Ippoliti2022fractal, Napp2022} to avoid mid-circuit measurement and access different manifestations of the underlying phases, from entanglement scaling~\cite{SkinnerPRX2019, LiPRB2018} to measurement-induced teleportation~\cite{Altman2021Teleportation}. We obtain finite-size signatures of a phase transition with a decoding protocol that correlates the experimental measurement with classical simulation data. The phases display sharply different sensitivity to noise, which we exploit to turn an inherent hardware limitation into a useful diagnostic. Our work demonstrates an approach to realize measurement-induced physics at scales that are at the limits of current NISQ processors.
\end{abstract} 

\maketitle 

The stochastic, non-unitary nature of measurement is a foundational principle in quantum theory and stands in stark contrast with the deterministic, unitary evolution prescribed by Schr\"{o}dinger's equation~\cite{Solvay2011}. Due to these unique properties, measurement is key to fundamental protocols in quantum information science such as teleportation~\cite{Bennett_teleportation_1993}, error correction~\cite{Shor_QEC_1995} and measurement-based computation~\cite{Raussendorf_MBQC_2003}.
All these protocols use quantum measurements, and classical processing of their outcomes, to build particular structures of quantum information in space-time.
Remarkably, such structures may also spontaneously emerge from random sequences of unitary interactions and measurements.
In particular, ``monitored'' circuits comprising both unitary gates and controlled projective measurements (Fig. 1\textbf{a}) were predicted to give rise to distinct nonequilibrium phases characterized by the structure of their entanglement~\cite{SkinnerPRX2019, LiPRB2018, Chan_unitary-projective_2019,Potter_review_2022,fisher2022random}---``volume-law''~\cite{Page_PRLentropy_1993} (extensive) or ``area-law''~\cite{Eisert_arealawRMP_2010} (limited) depending on the rate or strength of measurement. 

\begin{figure}[t!]
\centering
\includegraphics[width=0.48\textwidth]{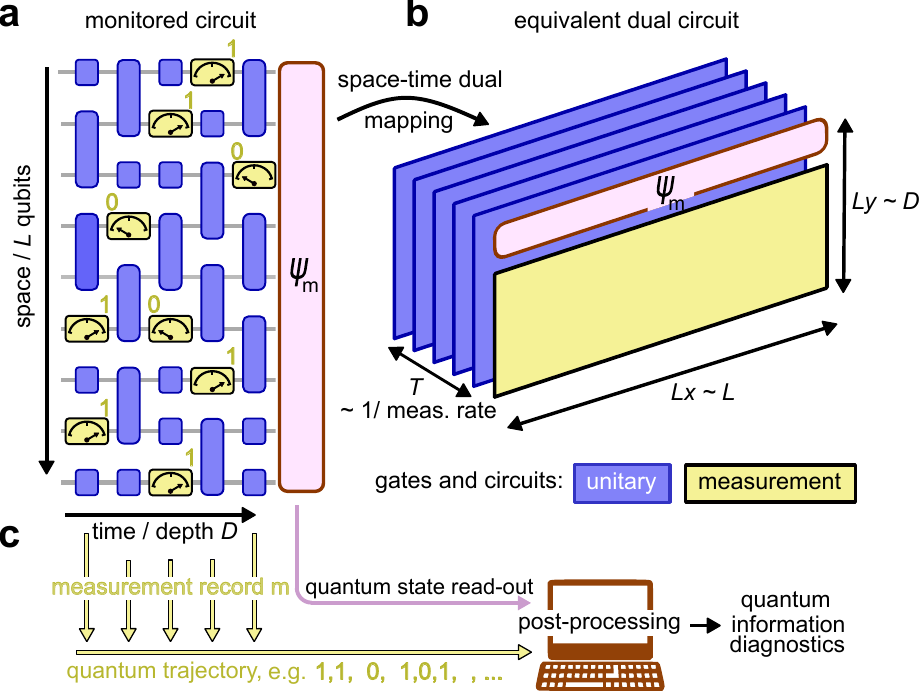}
\caption{\small{\textbf{Monitored circuits and space-time duality mapping. a,} A random $(1+1)$D quantum circuit composed of both unitary gates and measurements. \textbf{b,} 
An equivalent 2D shallow circuit of size $Lx \times Ly$ and depth $T$ with all measurements at the final time formed from a space-time duality mapping of the circuit in \textbf{a}. 
Due to the non-unitarity of measurements, there is freedom as to which dimensions are viewed as ``time'' and which as ``space''. In this example, $Ly$ is set by the $(1+1)$D circuit depth, $Lx$ by its spatial size, and $T$ is set by the measurement rate. \textbf{c,} Classical post-processing of the measurement record (quantum trajectory) and quantum state readout of a monitored circuit can be used to diagnose the underlying information structures in the system.}} \label{fig:fig1}
\end{figure}

\begin{figure*}[t!]
    \centering
    \includegraphics[width=0.98\textwidth]{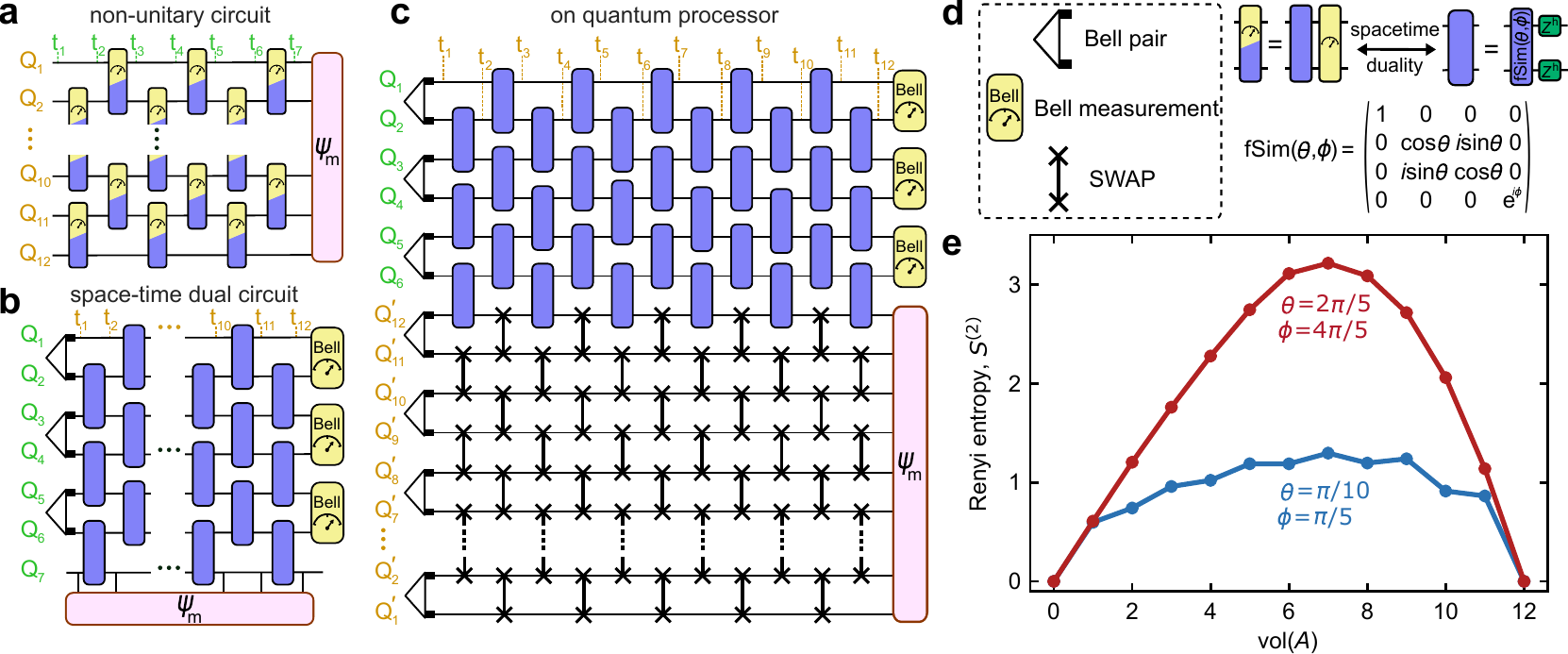} 
\caption{\small{\textbf{Implementation of space-time duality in 1D}. \textbf{a}, A quantum circuit composed of non-unitary 2-qubit operations in a brickwork pattern on a chain of 12 qubits with 7 time steps. Each 2-qubit operation can be a combination of unitary and measurement. \textbf{b}, The space-time dual of the circuit shown in \textbf{a} with the roles of space and time interchanged. The 12 qubit wave-function $|\psi_m \rangle$ is temporally extended along $Q_7$. \textbf{c}, In the experiment, a set of 12 ancilla qubits ($Q'_1$, ..., $Q'_{12}$) and a network of SWAP gates are used to teleport $|\psi_m \rangle $ onto the ancilla qubits. \textbf{d}, Illustration of the 2-qubit gate composed of a fSim unitary and random Z rotations with its space-time dual which is composed of a mixture of unitary and measurement operations. The power \textit{h} of the Z rotation is random for every qubit and periodic with each cycle of the circuit. \textbf{e}, Second Renyi entropy as a function of the volume of a subsystem $A$ via randomized measurements and postselection on $Q_1$, $\dots Q_6$. The data shown is noise-mitigated by subtracting an entropy density matching the total system entropy. See SI for justification.  } }
\end{figure*}  

Quantum processors, in principle, allow full control of both unitary evolution and projective measurements (Fig 1\textbf{a}). However, despite their significance in quantum information science, the experimental study of measurement-induced entanglement phenomena~\cite{NoelNatPhys2021, Minnich2022experimental} has been limited to small system sizes or efficiently simulatable Clifford gates. Due to the stochastic nature of measurement, detection of such phenomena requires either exponentially-costly postselection of measurement outcomes or more sophisticated data-processing techniques. This is because the phenomena are only visible in properties of quantum trajectories; a na\"{i}ve averaging of experimental repetitions incoherently mixes trajectories with different measurement outcomes and fully washes out the nontrivial physics. Additionally, implementation of the model in Fig. 1\textbf{a} requires mid-circuit measurements which are often problematic on superconducting processors since the time to perform a measurement is a much larger fraction of typical coherence times as compared to 2-qubit unitary operations. 
Here we leverage \textit{space-time duality} mappings to avoid mid-circuit measurements and develop a diagnostic of the phases based on a hybrid quantum-classical order parameter (akin to the cross-entropy benchmark of Ref.~\cite{arute2019quantum}) to overcome the issue of postselection. 
A matter of practical importance is the stability of these quantum information phases to noise. While relatively little is known about the effect of noise on monitored systems~\cite{LiFisher_opensystem_2021, WeinsteinAltman_PRL2022, Liu_noisyKPZ_2022}, noise is generally expected to destabilize measurement-induced non-equilibrium phases. Nonetheless, at the accessible system sizes, we show that noise serves as an independent probe of the phases.
Leveraging these insights we realize and diagnose 
measurement-induced phases of quantum information on system sizes up to 70 qubits. 

The space-time duality approach~\cite{Napp2022, Ippoliti2021postselection, Ippoliti2022fractal, Lu2021spacetime} enables more experimentally convenient implementations of monitored circuits by leveraging the absence of causality in such dynamics. 
When conditioning on measurement outcomes, the ``arrow of time'' loses its unique role and becomes interchangeable with spatial dimensions, giving rise to a network of quantum information in space-time~\cite{Vasseur_randomTN_2019} that can be analyzed in multiple ways. 
As an example, we may map 1D monitored circuits (Fig. 1\textbf{a}) to 2D shallow {\it unitary} circuits with measurements only at the final step (Fig. 1\textbf{b}, see Supplementary Information (SI) and Ref.~\onlinecite{Napp2022}), thus addressing the experimental issue of mid-circuit measurement. 

We begin by focusing on a special class of 1D monitored circuits that can be mapped by space-time duality to 1D unitary circuits. These models are theoretically well-understood~\cite{Ippoliti2022fractal, Lu2021spacetime} and convenient to implement experimentally. For families of operations that are dual to unitary gates\,(see SI), the standard model of monitored dynamics\,\cite{LiPRB2018, SkinnerPRX2019}based on a brickwork circuit of unitary gates and measurements\,(Fig.\,2\textbf{a}) can be equivalently implemented as a unitary circuit upon exchanging the space and time directions (Fig.~2\textbf{b}), leaving measurements only at the end. The desired output state $\ket{\psi_m}$ is prepared on a temporal subsystem, i.e. at a fixed position over different times~\cite{chertkov2022holographic}. It can be accessed without mid-circuit measurements by using ancillary qubits initialized in Bell pairs ($Q'_1$, $\dots Q'_{12}$ in Fig.\,2\textbf{c}) and SWAP gates, which teleport $\ket{\psi_m}$ onto the ancillary qubits at the end of the circuit\,(Fig.\,2\textbf{c}). While the resulting circuit still features postselected measurements, their reduced number (relative to a generic model, Fig.~2\textbf{a}) makes it possible to obtain the entropy of larger systems, up to all 12 qubits $Q_1',\dots Q_{12}'$, in individual quantum trajectories.

\begin{figure*}[t!]
    \centering
    \includegraphics[width=0.98\textwidth]{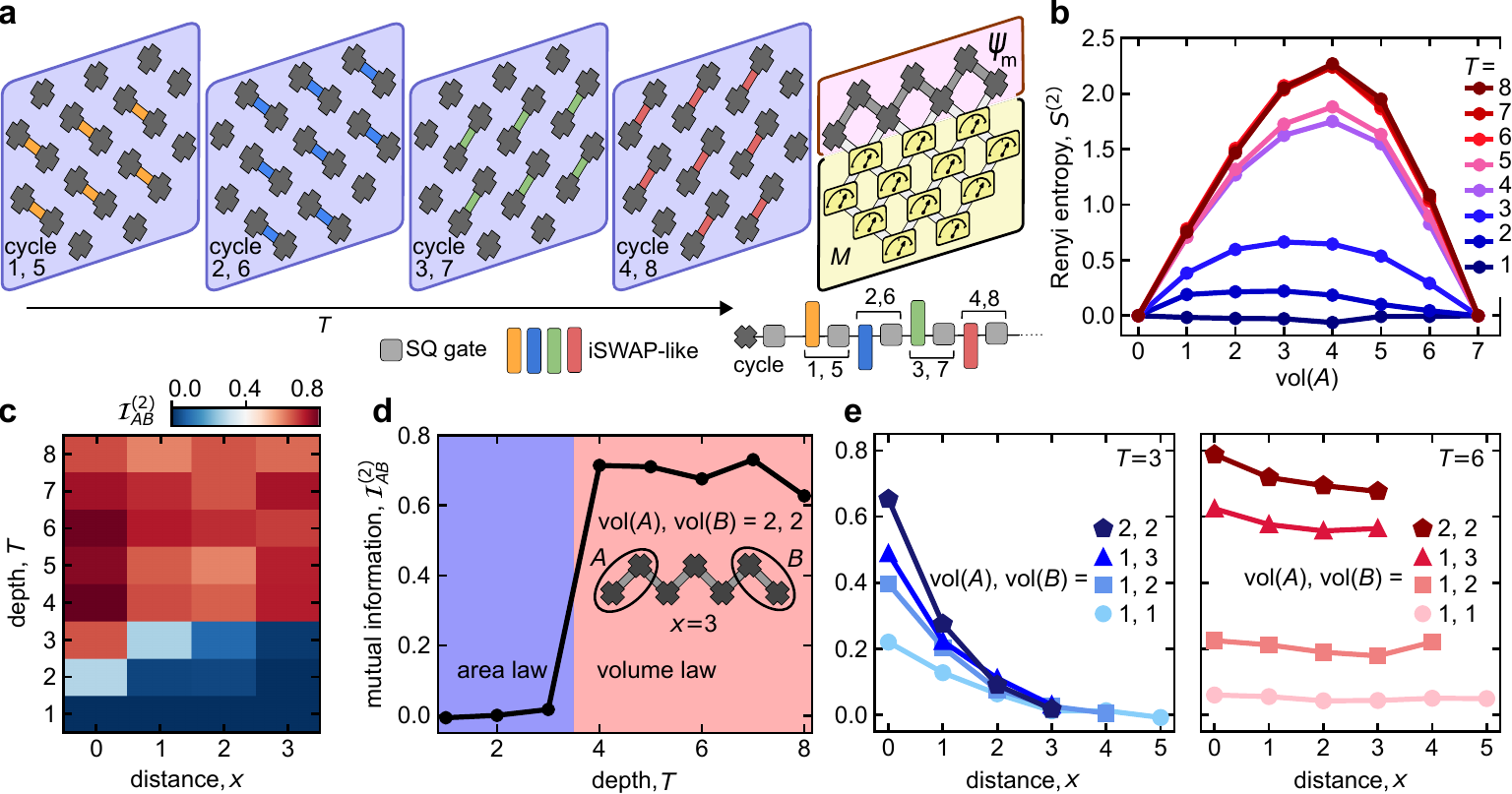} 
\caption{\small{\textbf{1D entanglement phases from 2D shallow quantum circuits}. \textbf{a}, Schematic of the 2D grid of qubits. At each cycle (blue boxes) of the circuit, random single qubit and two-qubit iSWAP-like gates are applied to each qubit in the cycle sequence shown. The random single qubit gate (grey) is chosen randomly from the set $\{\sqrt{X^{\pm1}}, \sqrt{Y^{\pm1}}, \sqrt{W^{\pm1}}, \sqrt{V^{\pm1}}\}$, where $W = (X+Y)/\sqrt{2}$ and $V = (X-Y)/\sqrt{2}$. At the end of the circuit, the lower $M=12$ qubits are measured and postselected on the most probable bitstring. \textbf{b}, Second Renyi entropy of contiguous subsystems $A$ of the $L=7$ edge qubits at various depths. The measurement is noise-mitigated in the same way as in Fig.\,2. \textbf{c}, second Renyi mutual information $\mathcal{I}_{AB}^{(2)}$ between 2-qubit subsystems $A$ and $B$ vs depth $T$ and distance $x$ (number of qubits between $A$ and $B$). \textbf{d}, $\mathcal{I}_{AB}^{(2)}$ as a function of $T$ for 2-qubit subsystems $A$ and $B$ at maximum separation. \textbf{e}, $\mathcal{I}_{AB}^{(2)}$ vs. $x$ for $T = 3$ and $T = 6$, respectively, for different volumes of $A$ and $B$.}}
\end{figure*} 

Prior works~\cite{Ippoliti2022fractal, Lu2021spacetime} predict distinct entanglement phases for $\ket{\psi_m}$ as a function of the choice of unitary gates in the dual circuit: volume-law entanglement if the gates induce an ergodic evolution and logarithmic entanglement if they induce a localized evolution. 
We implement unitary circuits representative of the two regimes, built from 2-qubit fermionic simulation (fSim) unitary  gates\,\cite{BrooksPRL2020} with swap angle $\theta$ and phase angle $\phi = 2\theta$, followed by random single qubit Z rotations. We choose angles $\theta = 2\pi/5$ and $\theta = \pi/10$ that are dual to non-unitary operations with different measurement strengths\,(Fig.\,2\textbf{d} and SI). 

To measure the second Renyi entropy on qubits composing $\ket{\psi_m}$, randomized measurements\,\cite{ZollerScience2019, Satzinger21} are performed on $Q'_1$, $\dots,Q'_{12}$. In Fig.\,2$\bold{e}$ we show the entanglement entropy as a function of subsystem size. 
The first gate set gives rise to a Page-like curve\,\cite{Page_PRLentropy_1993}, with entanglement entropy growing linearly in subsystem size up to half the system and then ramping down. The second instead shows a weak, sub-linear dependence of entanglement with subsystem size. These findings are consistent with the theoretical expectation of distinct entanglement phases (volume-law and logarithmic, respectively) in monitored circuits that are space-time dual to ergodic and localized unitary circuits~\cite{Ippoliti2022fractal, Lu2021spacetime}. A phase transition between the two can be realized by tuning the $(\theta,\phi)$ fSim gate angles.  

We next move beyond this specific class of circuits---with operations restricted to be dual to unitary gates---and investigate quantum information structures arising under more general conditions. Generic monitored circuits in 1D can be mapped onto shallow circuits in 2D, with final measurements on all but a 1D subsystem\,\cite{Napp2022}. The effective measurement rate $p$ is set by the depth of the shallow circuit $T$ and the number of measured qubits $M$. Heuristically, we have $p = M/ \,(M+L)T$ (number of measurements per unitary gate), where $L$ is the length of the chain of unmeasured qubits hosting the final state whose entanglement structure is investigated. Thus, for large $M$, a measurement-induced transition is tuned by varying $T$. 
We run 2D random quantum circuits~\cite{arute2019quantum} composed of iSWAP-like and random single qubit rotation unitaries on a grid of 19 qubits\,(Fig.\,3\textbf{a}), varying $T$ from 1 to 8. 
At each depth, we postselect on measurement outcomes of $M$ = 12 qubits and leave behind a 1D chain of $L$ = 7 qubits; its entanglement entropy is then measured for contiguous subsystems $A$ via randomized measurements. Two distinct behaviors are observed over a range of $T$ values\,(Fig.\,3\textbf{b}). For $T< 4$ the entropy scaling is sub-extensive with the size of the subsystem, while for $T \geq 4$ we observe an approximately linear scaling. 

\begin{figure*}[t!]
    \centering
    \includegraphics[width=0.97\textwidth]{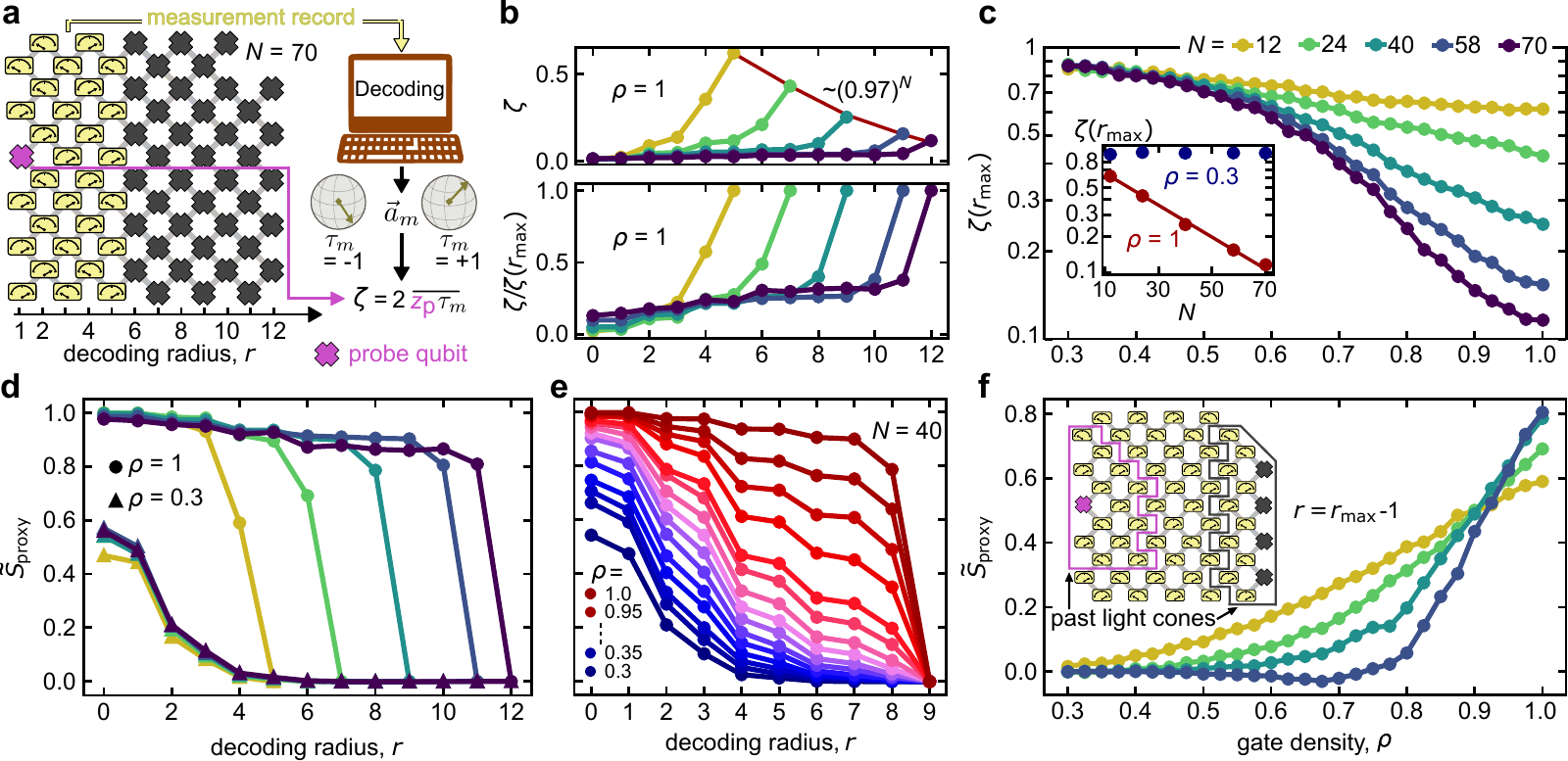} 
\caption{\small{\textbf{Decoding of local order parameter, measurement-induced teleportation and finite-size analysis}. \textbf{a}, schematic of the processor geometry and decoding procedure. The gate sequence is the same as in Fig. 3 with depth $T=5$. The decoding procedure involves classically computing the Bloch vector $\vec{a}_m$ of the probe qubit (pink) conditional on the experimental measurement record $m$ (yellow). The order parameter $\zeta$ is calculated via the cross-correlation between the measured probe bit $z_p$ and $\tau_m = {\sf sign}(\vec{a}_m\cdot\hat{z})$, which is $+1$ ($-1$) if $\vec{a}_m$ points above (below) the equator of the Bloch sphere. \textbf{b}, Decoded order parameter $\zeta$ and error-mitigated order parameter $\tilde{\zeta} = \zeta/\zeta(r_{\mathrm{max}})$ as a function of the decoding radius $r$ for different $N$ and $\rho = 1$. \textbf{c}, $\zeta(r_{\mathrm{max}})$ as a function of the gate density $\rho$ for different $N$. Inset shows that for small $\rho$, $\zeta(r_{\mathrm{max}})$ remains constant as a function of $N$ (disentangling phase), while for larger $\rho$, $\zeta(r_{\mathrm{max}})$ decays exponentially with $N$, implying sensitivity to noise of arbitrarily distant qubits (entangling phase). \textbf{d}, Error mitigated proxy entropy $\tilde{S}_{\mathrm{proxy}}$ as a function of the decoding radius for $\rho = 0.3$ (triangle) and  $\rho = 1$ (circle). In the disentangling phase, $\tilde{S}_{\mathrm{proxy}}$ decays rapidly to 0, independent of the system size. In the entangling phase, $\tilde{S}_{\mathrm{proxy}}$ remains large and finite up to $r_{\rm max}-1$.  \textbf{e}, $\tilde{S}_{\mathrm{proxy}}$ at $N = 40$ as a function of $r$ for different $\rho$, revealing a crossover between the entangling and disentangling phases for intermediate $\rho$. \textbf{f}, $\tilde{S}_{\mathrm{proxy}}$ at $r = r_{\rm max} - 1$ as a function of $\rho$ for $N = 12, 24, 40$, and $58$ qubits. Curves for different sizes approximately cross at $\rho_c \approx 0.9$. Inset schematically shows the decoding geometry for the experiment. Pink and grey lines encompass the past light cones (at depth $T=5$) of probe qubit and traced-out qubits at $r = r_{\rm max} - 1$, respectively. Data is collected over 2000 random circuit instances and 1000 shots each for every $N$ and $\rho.$}}
\end{figure*}

The spatial structure of quantum information can be further characterized by its signatures in correlations between disjoint subsystems of qubits: in the area-law phase entanglement decays rapidly with distance\,\cite{de2017quantum}, while in a volume-law phase, sufficiently large subsystems may be entangled arbitrarily far away. We study the second Renyi mutual information 
\begin{equation}
\mathcal{I}_{AB}^{(2)} = S_{A}^{(2)} + S_{B}^{(2)} - S_{AB}^{(2)},    
\end{equation}
\noindent between two subsystems $A$ and $B$ as a function of depth $T$ and the distance (number of qubits) $x$ between them (Fig.\,3\textbf{c}). For maximally-separated subsystems $A$ and $B$ of two qubits each, $\mathcal{I}_{AB}^{(2)}$ remains finite for $T \geq 4$, while it decays to 0 for $T \leq 3$ (Fig.\,3\textbf{d}). We also plot $\mathcal{I}_{AB}^{(2)}$ for subsystems $A$ and $B$ with several different sizes at $T=3$ and $T=6$ as a function of $x$ (Fig.\,3\textbf{e}). For $T = 3$ we observe a rapid decay of $\mathcal{I}_{AB}^{(2)}$ with $x$, indicating that only nearby qubits share information. For $T = 6$, however, $\mathcal{I}_{AB}^{(2)}$ does not decay with distance.

The observed structures of entanglement and mutual information provide strong evidence for the realization of measurement-induced area-law (``disentangling'') and volume-law (``entangling'') phases. Our results are indicative of a phase transition at critical depth $T\simeq 4$, consistent with previous numerical studies of similar models~\cite{Napp2022, Altman2021Teleportation, XiaoChen_2DshallowPRB_2022}. 
The same analysis without postselection on the $M$ qubits (see SI) shows vanishingly small mutual information, indicating that long-ranged correlations are in fact measurement-induced. 

The approaches we followed thus far are difficult to scale beyond system sizes of 10 to 20 qubits~\cite{Minnich2022experimental}, due to the exponential sampling complexity of postselecting measurement outcomes and obtaining entanglement entropy of extensive subsystems of the desired output states.
More scalable approaches have been recently proposed
\cite{GullansPRL2020,LiArXiv2022,Dehghani2022neural,Garratt_decoding_2022}
and implemented in efficiently-simulatable (Clifford) models~\cite{NoelNatPhys2021}. 
The key idea is that diagnostics of the entanglement structure must make use of both the read-out data from quantum state $\ket{\psi_m}$ and the classical measurement record $m$ in a classical post-processing step (Fig.~1\textbf{c}).
Postselection is the conceptually simplest instance of this idea: quantum read-out data is accepted or rejected conditional on $m$. 
However, as each instance of the experiment returns a random quantum trajectory~\cite{DalibardPRL92} out of $2^M$ ($M$ being the number of measurements), this approach incurs an exponential sampling cost which limits it to modest system sizes. Overcoming this issue ultimately requires more sample-efficient strategies which employ classical simulation~\cite{GullansPRL2020, Garratt_decoding_2022, LiArXiv2022}, possibly followed by active feedback~\cite{GullansPRL2020}. 

Here we develop a decoding protocol that correlates quantum read-out and measurement record to build a hybrid quantum-classical order parameter for the phases which is applicable to generic circuits and does not require active feedback on the quantum processor. 
A key idea is that the entanglement of a single ``probe'' qubit, conditioned on measurement outcomes, can serve as a proxy for the entanglement phase of the entire system\,\cite{GullansPRL2020}. This immediately eliminates one of the scalability issues---measuring entropy of extensive subsystems. 
The other issue---sampling of measurement outcomes---is handled by means of a classical simulation step, which allows us to make use of all experimental shots and is thus sample-efficient, unlike postselection. 

The protocol is illustrated in Fig.~4\textbf{a}. Each run of the circuit terminates with measurements returning binary outcomes $\pm 1$ for the probe qubit, $z_p$, and its surrounding $M$ qubits, $m$. The probe qubit is on the same footing as all others, and chosen at the post-processing stage. 
For each run, we classically compute the Bloch vector of the probe qubit conditional on the measurement record $m$, $\vec{a}_m$ (see SI). We then define $\tau_m = {\sf sign}(\vec{a}_m\cdot\hat{z})$, i.e. $+1$ if $\vec{a}_m$ points above the equator of the Bloch sphere, $-1$ otherwise. The cross-correlator between $z_p$ and $\tau_m$, averaged over many runs of the experiment such that the direction of $\vec{a}_m$ is randomized, yields an estimate of the length of the Bloch vector, $\zeta \simeq \overline{|\vec{a}_m|}$, which can in turn be used to define a proxy for the probe's entropy:

\begin{equation}
 \zeta = 2 \overline{z_p\,\tau_m}, \hspace{4mm} S_{\text{proxy}} = -\log_2[(1+\zeta^2)/2], 
\end{equation}
\noindent where over-line denotes averaging over all experimental shots and random circuit instances. A maximally-entangled probe corresponds to $\zeta = 0$.

In the standard teleportation protocol~\cite{Bennett_teleportation_1993}, a correcting operation conditional on the measurement outcome must be applied to retrieve the teleported state; in our decoding protocol, $\tau_m$ plays the role of the correcting operation, restricted to a classical bit-flip, and the cross-correlator describes the teleportation fidelity.  In the circuits relevant to our experiment (depth $T=5$ on $N\leq 70$ qubits), the classical simulation for decoding is tractable exactly. For arbitrarily large circuits, the existence of efficient decoders remains an open problem~\cite{GullansPRL2020, Dehghani2022neural, Barratt_PRLdecoding_2022}. Approximate decoders that work efficiently in part of the phase diagram, or for special models, also exist~\cite{GullansPRL2020}; we implement one such example based on matrix product states (see SI). 

We apply this decoding method to 2D shallow circuits acting on various subsets of a 70-qubit processor, consisting of $N\,=\,12, 24, 40, 58,$ and $70$ qubits in approximately square geometries\,(see SI). We choose a qubit near the middle of one side as the probe and compute the order parameter $\zeta$ by decoding measurement outcomes up to $r$ lattice steps away from that side while tracing out all others (Fig.\,4\textbf{a}). We refer to $r$ as the {\it decoding radius}. 
Due to measurements, the probe may remain entangled even as $r$ extends past its unitary light cone, corresponding to an emergent form of teleportation~\cite{Altman2021Teleportation}. 

As seen in Fig.\,3, the entanglement transition occurs as a function of depth $T$, with $3<T_c<4$. Because $T$ is a discrete parameter, it cannot be tuned to finely resolve the transition. To do so, we fix $T = 5$ and instead tune the density of gates, i.e., each iSWAP-like gate acts with probability $\rho$ and is skipped otherwise, setting an ``effective depth'' $T_{\rm eff} = \rho\,T$ that can be tuned continuously across the transition. Results for $\zeta(r)$ at $\rho=1$\,(Fig.\,4\textbf{b}) show a decay with system size $N$ of $\zeta(r_{\rm max})$, where $r=r_{\rm max}$ corresponds to measuring {all} qubits other than the probe. Such decay is purely due to noise in the system. 

Remarkably, sensitivity to noise can itself serve as an order parameter for the phases: in the disentangling phase, the probe is only affected by noise within a finite correlation length, while in the entangling phase it becomes sensitive to noise anywhere in the system. 
$\zeta(r_{\rm max})$ is shown as a function of $\rho$ for several $N$ values\,(Fig.~4\textbf{c}), indicating a transition at $\rho_c\approx 0.6-0.8$. At $\rho=0.3$, well below the transition, $\zeta(r_{\rm max})$ remains constant as $N$ increases (inset). In contrast, at $\rho = 1$ we fit $\zeta(r_{\rm max}) \sim 0.97^{N}$, indicative of an error rate $\sim 3\%$ per qubit for the entire sequence. This is approximately consistent with expectations for a depth 5 circuit based on individual gate and measurement errors rates\,(see SI). 
This response to noise is analogous to the susceptibility of magnetic phases to a symmetry-breaking field~\cite{JianPRB2020,LiVijay_DPRE_2021,WeinsteinAltman_PRL2022,Liu_noisyKPZ_2022} and thus sharply distinguishes the phases only in the limit of infinitesimal noise. For finite noise, we expect the $N$-dependence to be cut off at a finite correlation length.
We do not see the effects of this cut-off at the system sizes accessible to our experiment.

As a complementary approach, the underlying noiseless behavior may be estimated via noise mitigation. To this end we define the normalized order parameter $\tilde{\zeta}(r) = \zeta(r) / \zeta(r_{\rm max})$ and proxy entropy $\tilde{S}_{\rm proxy}(r) = -\log_2[(1+\tilde{\zeta}(r)^2)/2]$.
Persistence of entanglement with increasing $r$, corresponding to measurement-induced teleportation~\cite{Altman2021Teleportation}, diagnoses the entangling phase.
Fig.~4\textbf{d} shows the noise-mitigated entropy for $\rho = 0.3$ and $\rho = 1$, showing a rapid, $N$-independent decay in the former and a plateau up to $r = r_{\rm max}-1$ in the latter. 
At fixed $N = 40$, $\tilde{S}_{\rm proxy}(r)$ displays a crossover between the two behaviors for intermediate $\rho$ (Fig.\,4\textbf{e}).

To more sharply resolve this crossover, we show $\tilde{S}_{\rm proxy}(r_{\rm max}-1)$ as a function of $\rho$ for $N = 12$ up to $58$ (Fig.\,4\textbf{e}). 
The accessible system sizes approximately cross at $\rho_c\approx 0.9$. An upward drift of the crossing points with increasing $N$ is visible, confirming the expected instability of the phases to noise in the infinite-system limit. Nonetheless, signatures of the ideal finite-size crossing (estimated to be $\rho_c \simeq 0.72$ from noiseless classical simulation, see SI) remain recognizable at the sizes and noise rate accessible in our experiment, although moved to larger $\rho_c$. 
A stable finite-size crossing would mean that the probe qubit remains robustly entangled with qubits on the opposite side of the system even as $N$ increases. This is a hallmark of the teleporting phase~\cite{Altman2021Teleportation}, in which quantum information (aided by classical communication) travels faster than the limits imposed by locality and causality of unitary dynamics. Indeed, without measurements, the probe qubit and remaining un-measured qubits are causally disconnected, with non-overlapping past lightcones~\cite{Mi_2021} (pink and grey lines; inset of Fig.\,4\textbf{e}).

Our work focuses on the essence of measurement-induced phases: the emergence of distinct quantum information structures in space-time. We leveraged space-time duality mappings to circumvent mid-circuit measurements, devised scalable decoding schemes based on a local probe of entanglement, and exploited hardware noise to study these phases on up to 70 superconducting qubits. Our findings highlights the practical limitations of NISQ processors imposed by finite coherence. By identifying an exponential suppression of the decoded signal in the number of qubits, our results indicate that increasing the size of qubit arrays may not be beneficial without corresponding reductions in noise rates. At current error rates, extrapolation of our results (at $\rho = 1$, $T = 5$) to an $N$-qubit fidelity of less than $1\%$ suggests that arrays of over $\sim 150$ qubits would become too entangled with their environment for any signatures of the ideal (closed-system) entanglement structure to be practically detectable in experiment. This suggests an upper limit on qubit array sizes of about $12\times 12$ for this type of experiment, beyond which improvements in system coherence are needed.

\subsection*{Acknowledgments:}
We acknowledge discussions with E. Altman,  Y. Bao,  M. Block, M. Gullans, Y. Li, and L. Susskind. M.I. and V.K. thank T. Rakovszky for previous collaboration. D. Bacon is a CIFAR Associate Fellow in the Quantum Information Science Program. M.I. is supported  in  part  by  the Gordon and Betty Moore Foundation's EPiQS Initiative through Grant GBMF8686. V.K. acknowledges support from the US Department of Energy, Office of Science, Basic Energy Sciences, under Early Career Award Nos. DE-SC0021111, the Alfred P. Sloan Foundation through a Sloan Research Fellowship, and the Packard Foundation through a Packard Fellowship in Science and Engineering. Numerical simulations were performed in part on Stanford Research Computing Center's Sherlock cluster. We acknowledge the hospitality of the Kavli Institute for Theoretical Physics at the University of California, Santa Barbara (supported by NSF Grant PHY-1748958). 

\vspace{2mm}
\subsection*{Author contribution:}
J.C.H., M.I., X.M., V.K., and P.R. designed the experiment. J.C.H implemented the experiment. J.C.H and E.R. collected experimental data. M.I. designed the decoding protocol and implemented the classical simulations. J.C.H., M.I., X.M., V.K., and P.R. discussed the project in progress, interpreted the results, and wrote the manuscript. J.C.H. and M.I. wrote the supplementary material. X.M., V.K., and P.R. led and coordinated the project. Infrastructure support was provided by Google Quantum AI. All authors contributed to revising the manuscript and the supplement.

\vspace{2mm}
\subsection*{Data availability:}
The data that support the findings in this study are available at https://doi.org/10.5281/zenodo.7949563.

\onecolumngrid
\vspace{1em}
\begin{flushleft}
    {\hypertarget{authorlist}{${}^\dagger$}  \small Google Quantum AI and Collaborators}

    \bigskip

    \renewcommand{\author}[2]{#1\textsuperscript{\textrm{\scriptsize #2}}}
    \renewcommand{\affiliation}[2]{\textsuperscript{\textrm{\scriptsize #1} #2} \\}
    \newcommand{\corrauthora}[2]{#1$^{\textrm{\scriptsize #2}, \hyperlink{corra}{*}}$}
    \newcommand{\corrauthorb}[2]{#1$^{\textrm{\scriptsize #2}, \hyperlink{corrb}{\mathsection}}$}
    \newcommand{\xGoogle}{\affiliation{1}{Google Research, Mountain View, CA, USA}}

\newcommand{\xStanford}{\affiliation{2}{Department of Physics, Stanford University, Stanford, CA, USA}}

\newcommand{\xTexas}{\affiliation{3}{Department of Physics, University of Texas at Austin, Austin, TX, USA}}

\newcommand{\xCornell}{\affiliation{4}{Department of Physics, Cornell University, Ithaca, NY, USA}}

\newcommand{\xUMass}{\affiliation{5}{Department of Electrical and Computer Engineering, University of Massachusetts, Amherst, MA, USA}}

\newcommand{\xUCONN}{\affiliation{6}{Department of Physics, University of Connecticut, Storrs, CT}}

\newcommand{\xAU}{\affiliation{7}{Department of Electrical and Computer Engineering, Auburn University, Auburn, AL, USA}}

\newcommand{\xCQC}{\affiliation{8}{QSI, Faculty of Engineering \& Information Technology, University of Technology Sydney, NSW, Australia}}

\newcommand{\xUCR}{\affiliation{9}{Department of Electrical and Computer Engineering, University of California, Riverside, CA, USA}}

\newcommand{\xCU}{\affiliation{10}{Department of Chemistry, Columbia University, New York, NY, USA}}

\newcommand{\xUoCA}{\affiliation{11}{Department of Physics and Astronomy, University of California, Riverside, CA, USA}}

\begin{footnotesize}

\newcommand{\Google}{1}
\newcommand{\Stanford}{2}
\newcommand{\Texas}{3}
\newcommand{\Cornell}{4}
\newcommand{\UMass}{5}
\newcommand{\UCONN}{6}
\newcommand{\AU}{7}
\newcommand{\CQC}{8}
\newcommand{\UCR}{9}
\newcommand{\CU}{10}
\newcommand{\UoCA}{11}

\corrauthora{J. C. Hoke}{\Google,\! \Stanford},
\corrauthora{M. Ippoliti}{\Stanford,\! \Texas},
\author{E. Rosenberg}{\Google,\! \Cornell},
\author{D. Abanin}{\Google},
\author{R. Acharya}{\Google},
\author{T. I. Andersen}{\Google},
\author{M. Ansmann}{\Google},
\author{F. Arute}{\Google},
\author{K. Arya}{\Google},
\author{A. Asfaw}{\Google},
\author{J. Atalaya}{\Google},
\author{J. C.~Bardin}{\Google,\! \UMass},
% ...

\bigskip

\xGoogle
\xStanford
\xTexas
\xCornell
\xUMass
\xUCONN
\xAU
\xCQC
\xUCR
\xCU
\xUoCA



{\hypertarget{corra}{$*$} These authors contributed equally to this work.}



\end{footnotesize}
\end{flushleft}

\newpage
\twocolumngrid
\bibliography{References.bib}
\end{document}


\title{Supplement - Measurement-induced entanglement \\ and teleportation on a noisy quantum processor}
\author{Google Quantum AI and Collaborators}
\date{\today}

\maketitle
\tableofcontents

\newpage
    
\section{Quantum processor details}\label{gatecalib}
\subsection{Coherence times and readout error}\label{coherence_times}

\begin{figure*}[t!]
    \centering
    \includegraphics[width=0.7\textwidth]{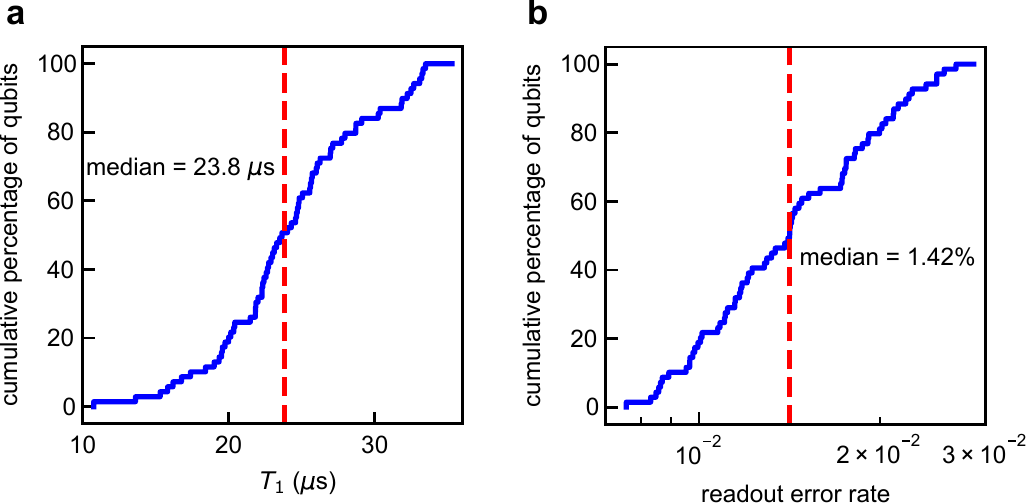} 
\caption{\small{\textbf{a}, Cumulative distribution of $T_{1}$ across the 70 qubit superconducting quantum processor with a dashed line indicating a median value of 23.8 $\mu$s. \textbf{b}, Cumulative distribution of the readout error across the 70 qubit superconducting quantum processor with a dashed line indicating the median value of 1.42\%.} \label{fig:coherence_readout}}
\end{figure*}

The experiment is performed on a quantum processor with 70 Transmon qubits with tunable frequencies and interqubit couplings with a similar design to Ref. \cite{arute2019quantum}. In Fig.~\ref{fig:coherence_readout}\textbf{a} we show the characteristic relaxation ($T_{1}$) times of the entire 70 qubit chip with a median of 23.8 $\mu$s, as measured through simultaneous population decay ($\ket{1} \rightarrow \ket{0}$) experiments.

We also benchmark the readout errors of the entire 70 qubit chip, which are especially important to minimize to study measurement-induced physics. All qubits are repeatedly and simultaneously prepared in a random bit string state and then readout. For each qubit, an error occurs if the resulting bit does not match its initial state. We achieve a median single qubit readout error rate of 1.42\% (Fig.~\ref{fig:coherence_readout}\textbf{b}).

\subsection{Single and two qubit errors}\label{readout_and_gate_errors}
\begin{figure*}[t!]
    \centering
    \includegraphics[width=0.7\textwidth]{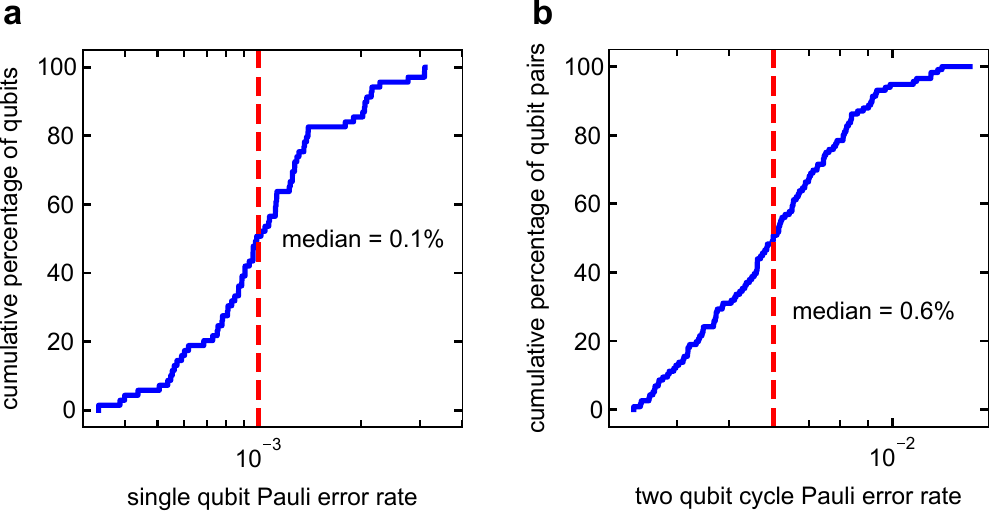} 
\caption{\small{\textbf{a}, \textbf{b}, Cumulative distributions of the single-qubit Pauli error (\textbf{a}) and two-qubit cycle iSWAP-like Pauli error (\textbf{b}) across the 70 qubit superconducting quantum processor as measured by XEB. We find median values of 0.10\% and 0.60\% for the single-qubit and iSWAP-like total cycle errors, respectively.} \label{fig:gate_errors}}
\end{figure*}

We benchmark the error rates of the single qubit and two qubit iSWAP-like gates used in the experiment using cross entropy benchmarking (XEB)\cite{arute2019quantum}. In Fig.~\ref{fig:gate_errors} we show the cumulative distributions of single-qubit and two qubit cycle Pauli error rates, with medians of 0.10\% and 0.60\%, respectively. The error of a cycle corresponds to the error of two randomly chosen single-qubit gate on each qubit followed by the iSWAP-like gate.

\section{2-Qubit fSim gates}\label{sec2}

In the following sections we describe the implementation of arbitrary 2-qubit fSim gates with superconducting qubits. We then characterize the leakage errors that arise from these interactions. Finally, we discuss the specifics related to the calibration of the different fSim gates used in this experiment and their resulting fidelities.

\subsection{fSim implementation}\label{fsim implementation}

\begin{figure*}[t!]
    \centering
    \includegraphics[width=0.65\textwidth]{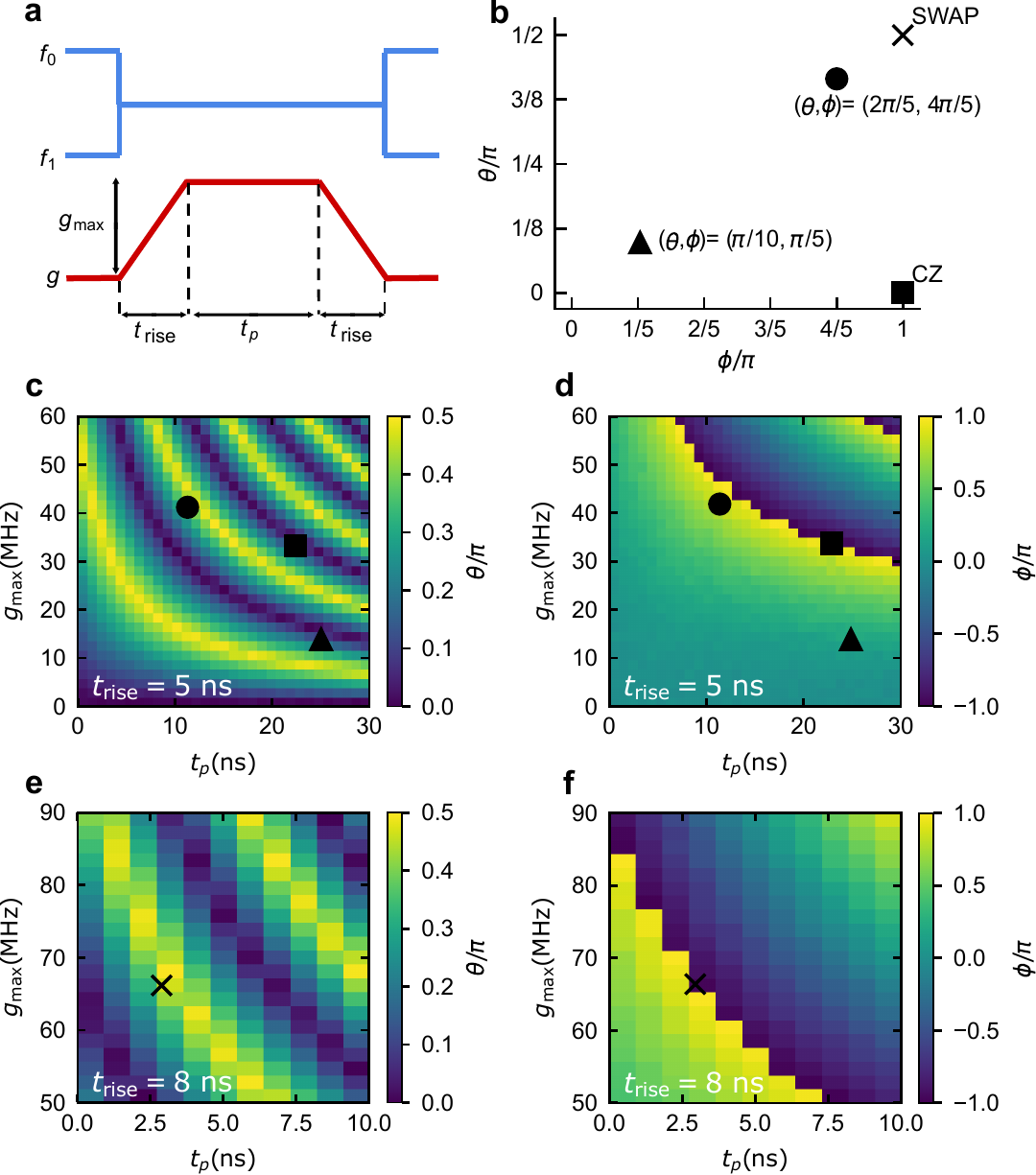} 
\caption{\small{\textbf{a}, Schematic illustration of the DC pulse used to implement arbitrary fSim gates. \textbf{b,} Schematic plot showing the four sets of SWAP and conditional phase angles to realize the different fSim gates used in our experiment: 
 ($\theta$, $\phi$) = ($2\pi/5$, $4\pi/5$) (circle), ($\pi/10$, $\pi/5$) (triangle), ($0$, $\pi$) (CZ, square), and ($\pi/2$, $\pi$) (SWAP, cross). \textbf{c, d,} Experimentally measured values of $\theta$ and $\phi$ as a function of the hold time $t_{p}$ and maximum coupling strength $g_{\mathrm{max}}$ for $t_{\mathrm{rise}} = 5$ ns. The approximate pulse parameters for the fSim($2\pi/5$, $4\pi/5$), fSim($\pi/10$, $\pi/5$), and CZ gates are overlaid on top. \textbf{e, f,} Same as \textbf{c, d,} but for $t_{\mathrm{rise}} = 8$ ns. The approximate pulse parameters for the SWAP gate is overlaid on top.}\label{fig:fsim_ut}}
\end{figure*}

All of two qubit gates used in the 1D experiment highlighted in Fig. 2 in the main text are implemented as general fermionic simulation unitaries of the form

\begin{equation}
\mathrm{fSim}(\theta, \phi, \Delta_{+}, \Delta_{-}, \Delta_{-, \mathrm{off}}) =
\begin{pmatrix}
1 & 0 & 0 & 0\\
0 & e^{i(\Delta_{+} + \Delta_{-})} \mathrm{cos} \ \theta & -i e^{i(\Delta_{+} - \Delta_{-, \mathrm{off}})} \mathrm{sin} \ \theta & 0\\
0 & -i e^{i(\Delta_{+} + \Delta_{-, \mathrm{off}})} \mathrm{sin} \ \theta & e^{i(\Delta_{+} - \Delta_{-})} \mathrm{cos} \ \theta & 0\\
0 & 0 & 0 & e^{i(2\Delta_{+} - \phi)}
\end{pmatrix},
\end{equation}

where $\theta$ is the SWAP angle, $\phi$ is the conditional phase, and the $\Delta$'s are single qubit phases. The general pulse shape used to implement arbitrary fSim gates of this form is displayed in Fig.~\ref{fig:fsim_ut}\textbf{a}. The pulse shape is defined by the maximum coupling strength $g_{\mathrm{max}}$, the hold time at $g_{\mathrm{max}}$, $t_{p}$, and the rise time $t_{\mathrm{rise}}$. During the interaction the fundamental frequencies $f_0$ and $f_1$ of the two qubits are brought into resonance. During the time the coupler is on ($2 t_{\mathrm{rise}} + t_{p}$), resonant interactions between $\ket{01}$ and $\ket{10}$ of the two qubits lead to a finite value of $\theta$, while dispersive interactions between $\ket{11}$ and $\ket{02}$ (and $\ket{20}$) lead to a finite $\phi$. In terms of the coupler pulse parameters, both $\theta$ and $\phi$ scale linearly with $t_{p}$ while $\theta \propto g_{\mathrm{max}}$ and $\phi \propto g^{2}_{\mathrm{max}}$. This difference in scaling enables independent control of $\theta$ and $\phi$. The single qubit phases occur due to the frequency detunings of qubits during the DC pulse and are typically calibrated and set to near zero. We show typical maps of experimentally measured $\theta$ and $\phi$ via unitary tomography as a function of $t_{p}$ and $g_{\mathrm{max}}$, with $t_{\mathrm{rise}} = 5$ ns (Fig.~\ref{fig:fsim_ut}\textbf{c, d}) and $8$ ns (Fig.~\ref{fig:fsim_ut}\textbf{e, f}). This implementation of the fSim gate is built off of the technique introduced in Ref.\cite{Morvan_2022}, where further details can be found.

\subsection{Characterizing leakage errors}\label{leakage errors}

\begin{figure*}[t!]
    \centering
    \includegraphics[width=0.8\textwidth]{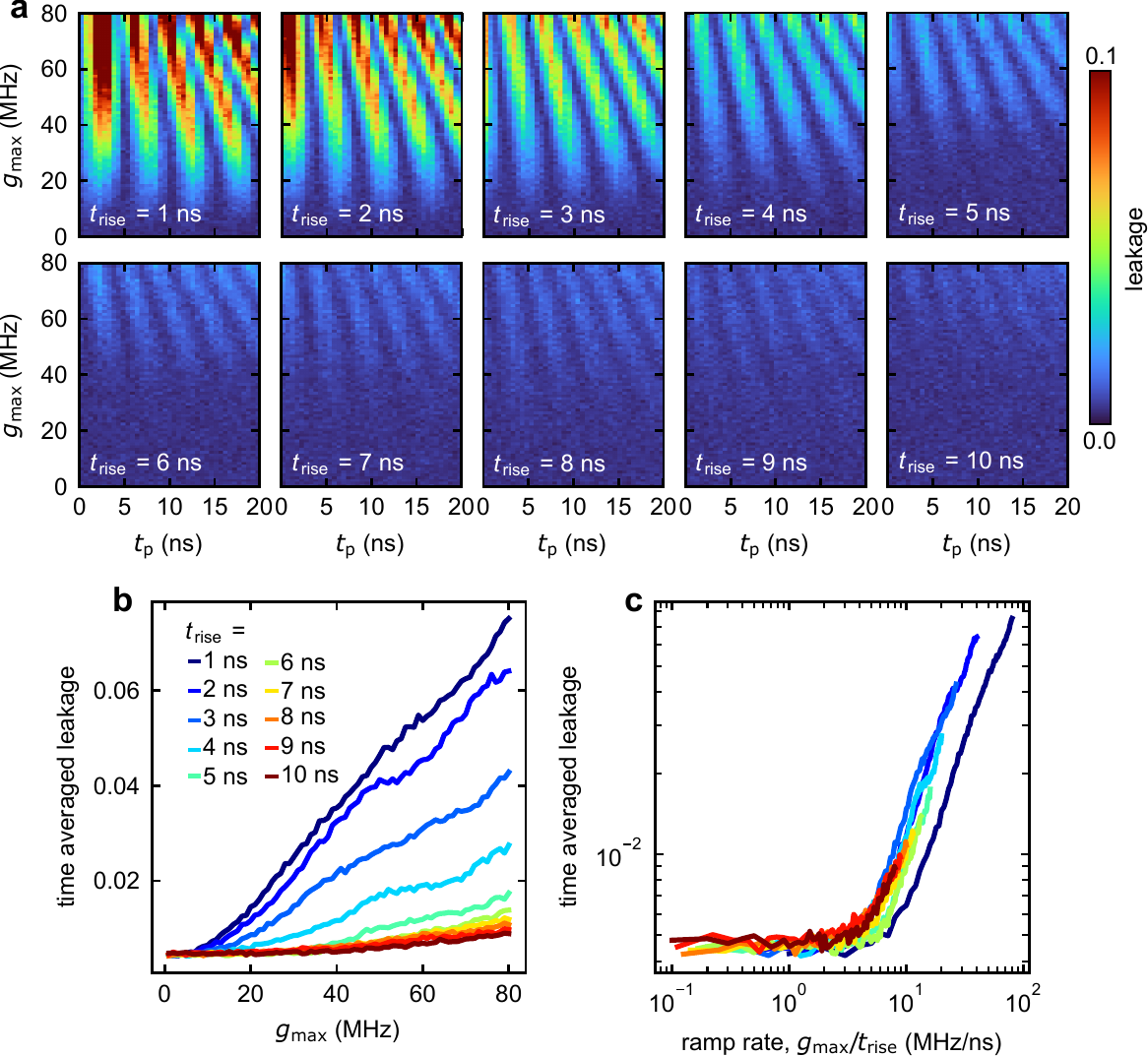} 
\caption{\small{\textbf{a}, Leakage measured as a function of the coupling strength $g_{\mathrm{max}}$ and hold time $t_{p}$ for fixed values of the rise time $t_{\mathrm{rise}}$. \textbf{b,} The time averaged leakage (averaged over the range $0 < t_{p} < 20$ ns) as a function of $g_{\mathrm{max}}$ for different values of $t_{\mathrm{rise}}$. \textbf{c, } The time averaged leakage plotted as a function of the ramp rate $g_{\mathrm{max}}/t_{\mathrm{rise}}$ for different $t_{\mathrm{rise}}$.}\label{fig:leakage}}
\end{figure*}

For an arbitary fSim gate, the above mentioned interactions with states outside the computational subspace naturally lead to leakage errors. Due to decoherence effects it is desirable to implement fast coupler pulses, while longer ones are favorable for suppressing leakage error. Furthermore, leakage become more important sources of errors as $g_{\mathrm{max}}$ increases (Fig.~\ref{fig:leakage}). Although the leakage oscillates as a function of $t_{p}$ and one could carefully select the coupler gate's pulse parameters to sit at one of the leakage minima, this does not allow for precise control of the resulting SWAP angle $\theta$ and conditional phase $\phi$; if the desired $(\theta, \phi)$ happen to lie on top of a leakage maxima, the resulting fidelity of the gate will suffer. Therefore if precise control of $(\theta, \phi)$ is needed, an alternative strategy is required to minimize leakage, particularly at large $g_{\mathrm{max}}$.

In Ref. \cite{Morvan_2022} leakage errors were minimized by implementing a trapezoidal coupler pulse with $t_{\mathrm{rise}}$ set to 5 ns. Indeed this was sufficient for realizing relatively high fidelity (99\%) fSim gates for a select few $(\theta, \phi)$, however further characterizations are needed for arbitrary $(\theta, \phi)$. 
To investigate the effect of the rise time on leakage rates, we performed a systematic study of leakage as a function of $g_{\mathrm{max}}$, $t_{p}$, and $t_{\mathrm{rise}}$. To measure leakage, qubit pairs are initialized in the $\ket{11}$ state, allowed to interact via the fSim unitary defined by the pulse sequence in Fig.~\ref{fig:fsim_ut}\textbf{a}, and then read out. Here, we measure leakage as the probability of finding the qubits in one of the leakage states ($\ket{02}$ or $\ket{20}$) after interacting via the fSim pulse, similar to Ref. \cite{BrooksPRL2020}. In Fig.~\ref{fig:leakage}\textbf{a} we show maps of the leakage at fixed values of $t_{\mathrm{rise}}$ for $0 < g_{\mathrm{max}} < 80$ MHz and $0 < t_{p} < 20$ ns. As can be seen, leakage is generally suppressed as $t_{\mathrm{rise}}$ increases, which sets the ramp rate of the coupler pulse for a given $g_{\mathrm{max}}$ and the degree of ``adiabaticity" of the gate. This can also be seen by plotting the leakage, averaged over $0 < t_{p} < 20$ ns, as a function of $g_{\mathrm{max}}$ for various $t_{\mathrm{rise}}$ (Fig.~\ref{fig:leakage}\textbf{b}). The fact that the leakage depends predominately on the ramp rate $g_{\mathrm{max}}/t_{\mathrm{rise}}$, rather than the set rise time, is illustrated in Fig.~\ref{fig:leakage}\textbf{c}, where the time averaged leakage for each $t_{\mathrm{rise}}$ collapses onto the same curve as a function of the ramp rate. While this eliminates the knowledge that leakage indeed oscillates as a function of $t_{p}$, it highlights a crossover from an approximately ramp rate-independent leakage behavior to one in which the leakage increases approximately proportionally with the ramp rate.

\subsection{fSim gate calibration}\label{fsim gate calibration}

\begin{figure*}[t!]
    \centering
    \includegraphics[width=0.6\textwidth]{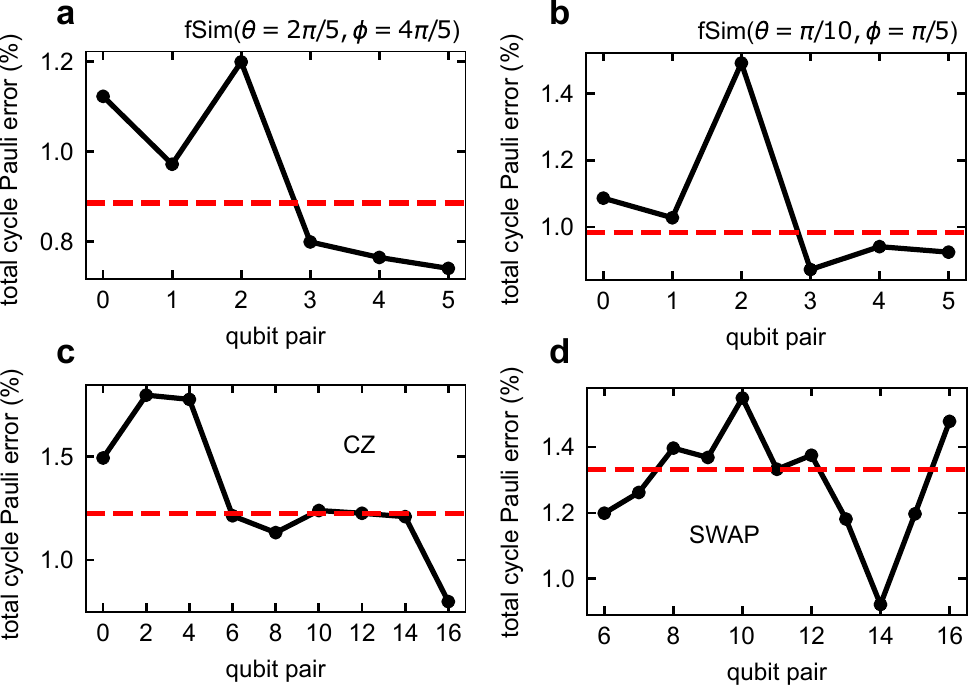} 
\caption{\small{\textbf{a-d}, Cycle Pauli error of the fSim($2\pi/5$, $4\pi/5$), fSim($\pi/10$, $\pi/5$), CZ, and SWAP gates, respectively, as measured by cross entropy benchmarking (XEB) for the qubit pairs that utilized each gate. Red dashed line corresponds to the median.}\label{fig:fsim_xeb}}
\end{figure*}

In our experiment we implemented four different fSim gates: fSim($\theta=2\pi/5$, $\phi=4\pi/5$), fSim($\theta=\pi/10$, $\phi=\pi/5$), CZ = fSim($\theta=0$, $\phi=\pi$), and SWAP = fSim($\theta=\pi/2$, $\phi=\pi$, $\Delta_{+}=\pi/2$), where all of the $\Delta$ angles are set to 0 unless explicitly stated otherwise. 
In previous works the SWAP was implemented by concatenating several single qubit and two qubit gates in series~\cite{Harrigan_2021, nonabelian_2022, wormhole_2022}. While this works in theory, the SWAP as a composite gate results in poor error rates and a resulting long gate duration, which is undesirable for decoherence effects. A better, alternative approach is to implement SWAP with a single coupler pulse using our fSim framework. Following the analysis of the leakage, we found that $t_{\mathrm{rise}}$ = 8 ns was sufficient for minimizing its leakage at large $g_{\mathrm{max}}$, and accordingly set $t_{\mathrm{rise}}$ = 8 ns for the calibration of the SWAP gate, while $t_{\mathrm{rise}}$ = 5 ns was set for the others. The approximate coordinates of the four fSim gates are overlaid on top of the ($\theta$, $\phi$) maps in Fig.~\ref{fig:fsim_ut}. The target angles are calibrated by running higher resolution, local  unitary tomography maps $\theta(t_{p}, g_{\mathrm{max}})$ and $\phi(t_{p}, g_{\mathrm{max}})$ near an initial guess based on the maps shown in Fig.~\ref{fig:fsim_ut}. The optimal pulse parameters for target angles ($\theta_{0}$, $\phi_{0}$) are then found by minimizing the cost function 
\begin{equation}
f(\theta_{0}, \phi_{0}, t_{p}, g_{\mathrm{max}} ) = \mathrm{log}[(\theta_{0} - \theta(t_{p}, g_{\mathrm{max}}))^2 +  (\phi_{0} - \phi(t_{p}, g_{\mathrm{max}}))^2].
\end{equation}
The remaining single-qubit angles are set to 0, except for SWAP where $\Delta_{+} = \pi/2$. The resulting total cycle Pauli error rates of the qubit pairs that utilized each fSim gate for the experiment highlighted in Fig. 2 of the main text are displayed in Fig.~\ref{fig:fsim_xeb}. We achieve median cycle Pauli error rates of 0.89\%, 0.98\%, 1.33\%, and 1.23\% for fSim($2\pi/5$, $4\pi/5$), fSim($\pi/10$, $\pi/5$), CZ, and SWAP gates, respectively as measured by XEB. 

\newpage

\section{Measuring entropy via randomized measurements}\label{sec3}

\subsection{Randomized measurements}\label{randomized_measurements}
In our experiments we utilize the randomized measurement protocol \cite{van_Enk_2012, Elben_2018, ZollerScience2019, Satzinger21} to measure the second Renyi entropy. For $N$ qubits, the protocol relies on applying random single qubit unitaries $U = u_{1} \otimes u_{2} \otimes ... \otimes u_{N}$, each independently drawn from the circular unitary ensemble (CUE). For a system $A$ of $N_{A} \le N$ qubits, the purity is given by
\begin{equation}
\mathrm{Tr}(\rho_{A}^{2}) = 2^{N_{A}} \sum_{s, s^{'}} (-2)^{-H(s, s')} \langle P(s)P(s') \rangle_{U},
\end{equation}
where $\rho_{A}$ is the density matrix of $A$, $H(s, s')$ is the Hamming distance between binary bitstrings $s$ and $s'$ in the computational basis and $P(s)$ is the probability of measuring $s$. The average over probabilities of bitstrings $\langle P(s)P(s') \rangle_{U}$ is averaged over several instances of $U$ and in practice is estimated with jackknife resampling to remove bias. The second Renyi entropy is then given by $S^{(2)}_{A} = -\mathrm{log}_{2}(\mathrm{Tr}(\rho_{A}^{2}))$. 
A particular advantage of the randomized measurement method is that the same set of measurements can be used to simultaneously calculate the entropies for multiple subsystems. This is especially beneficial for calculating the second Renyi mutual information ${I}_{AB}^{(2)} = S_{A}^{(2)} + S_{B}^{(2)} - S_{AB}^{(2)}$.
Previous successful implementations of the technique can be found in Refs. \cite{ZollerScience2019, Satzinger21}.

\subsection{Entropy error mitigation and additional data and numerics}\label{entropy error mitigation}

In today's quantum processors noise is pervasive and unavoidable. Since noise adds a background entropy, it can lead to important deviations from theoretically predicted values. Consequently, it is essential to account for and mitigate the effects of noise when measuring entropy in an experiment. Our method for error mitigating noise relies on the fact that we perform randomized measurements of entropy on the entirety of a 1D chain that is expected to be in a pure state; any entropy of the global state of the chain must come solely from noise contributions. We further assume that background entropy due to noise is extensive and with a spatially-uniform density, such that for any subsystem $A$ of the pure state $AB$, the background entropy $\delta S$ is given by
\begin{equation}
\delta S(A) = \frac{\mathrm{vol}(A)}{\mathrm{vol}(AB)} \delta S(AB). \label{eq:supp_error_mitigation_def}
\end{equation}

\begin{figure*}[t!]
    \centering
    \includegraphics[width=0.7\textwidth]{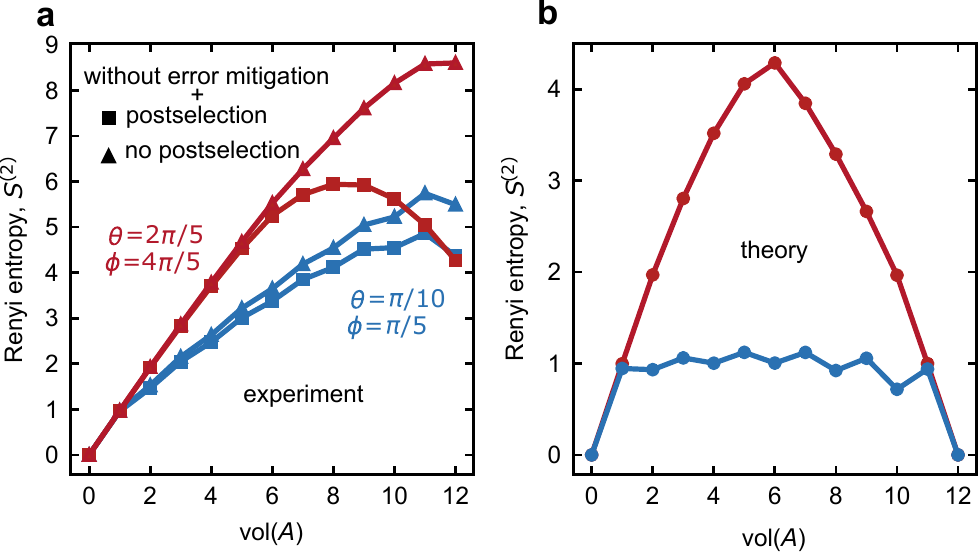} 
\caption{\small{\textbf{a}, Non error-mitigated Renyi entropy as a function of the volume of a subsystem $A$ when postselecting (squares) and not postselecting (triangles) on the measurement record of qubits $Q_1$, $Q_2$, $Q_3$, $Q_4$, $Q_5$, and $Q_6$ for the experiment implemented in Fig. 2 of the main text. 
\textbf{b}, Results of an exact numerical simulation of the second Renyi entropy as a function of the volume of a subsystem $A$ for the experiment implemented in Fig. 2 of the main text. The curves agree qualitatively well with the experimental results.}\label{fig:1d_entropy}}
\end{figure*}

In Fig.~\ref{fig:1d_entropy}\textbf{a} we show the raw, non-error-mitigated entropy curves for the two examples of a 18 qubit space-time dual circuit. For each curve, bitstrings are collected over 20 random instances (different random choices of single qubit gates) of the circuit with 5 instances of CUE unitaries $U$ each and 10 million shots per instance of $U$.  We then postselect on the most probable measurement outcome of the 6 measured qubits and use the outcome of the randomized measurements to extract the second Renyi entropy of subsystems of the 12 remaining qubits. Each data point corresponds to an average over all contiguous subsystems of the specified volume. There is a significant residual entropy for the state of the whole system due to noise, which grows both with the size of the chain and the depth of the circuit. 
We then apply our error mitigation technique by subtracting a straight line that intersects the origin and the whole-system ($\mathrm{vol}(A) = 12$) entropy. In Fig.~\ref{fig:1d_entropy}\textbf{b} we show exact numerical simulations of the second Renyi entropy, which show good qualitative agreement with the experiment.

\begin{figure*}[t!]
    \centering
    \includegraphics[width=0.9\textwidth]{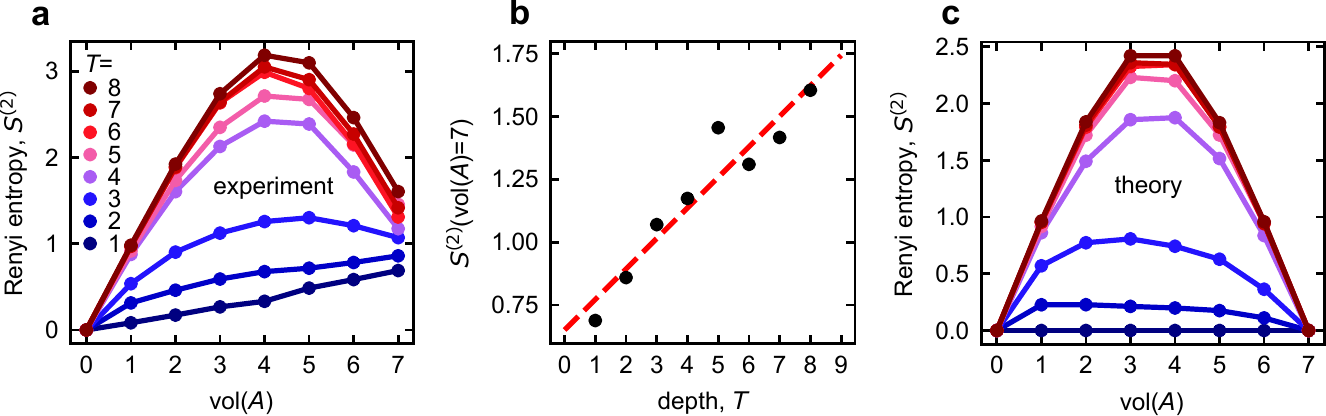} 
\caption{\small{\textbf{a}, Experimentally measured second Renyi entropy as a function of a sub-system $A$ of the chain of 7 qubits at various depths without error mitigation. \textbf{b}, Second Renyi entropy of the entire 7 qubit chain after measurement of the lower 12 qubits as a function of circuit depth. The entropy grows approximately linearly due to decoherence in the system. Dashed line is a linear fit and intercepts the $y$-axis at a non-zero value, which we attribute to readout error.  \textbf{c}, Exact numerical simulation of the second Renyi entropy as a function of the volume of a sub-system $A$ of the 7 edge qubits at the same depths as in \textbf{a}. The results agree qualitatively well with the error-mitigated data presented in the main text.}\label{fig:2d_entropy}}
\end{figure*}

We apply this same strategy for the randomized measurements of our experiment on a 2D grid of 19 qubits. In Fig.~\ref{fig:2d_entropy}\textbf{a} we show the experimental results without error mitigation as a function of the circuit depth $T$. For each curve, bitstrings are collected over 20 random instances (different random choices of single qubit gates) of the circuit with 5 instances of CUE unitaries $U$ each and 20 million shots per instance of $U$. We then postselect on the most probable measurement outcome of the 12 measured qubits and use the outcome of the randomized measurements to extract the second Renyi entropy of subsystems of the 7 remaining qubits. Each data point corresponds to an average over all contiguous subsystems of the specified volume. As expected, the whole-system entropy (vol($A$) = $7$) grows approximately linearly with circuit depth (Fig.~\ref{fig:2d_entropy}\textbf{b}). Noticeably, the whole-system entropy intersects the $y$-axis at a finite value of $\delta S$. We attribute this residual background entropy to readout error, which remains an approximately constant source of error as a function of circuit depth. We note that this residual entropy due to readout error comes both from error in the readout of the 12 postselected qubits and of the remaining chain of 7 qubits whose entropy we obtain via randomized measurements. The results for an exact numerical simulation (Fig.~\ref{fig:2d_entropy}\textbf{c}) agrees qualitatively with the experiment. 

\begin{figure*}[t!]
    \centering
    \includegraphics[width=0.35\textwidth]{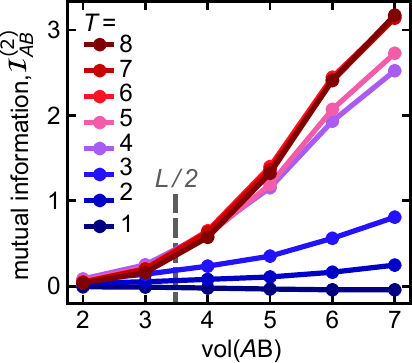} 
\caption{\small{Second Renyi mutual information $\mathcal{I}_{AB}^{(2)}$ as a function of the combined volume of $A$ and $B$ for each $T$. Each point is averaged over all possible configurations of contiguous subsystems $A$ and $B$ for a fixed total volume (i.e. averaged over all values of $x$ between $A$ and $B$). }\label{fig:qmi_ab}}
\end{figure*}

\begin{figure*}[t!]
    \centering
    \includegraphics[width=0.94\textwidth]{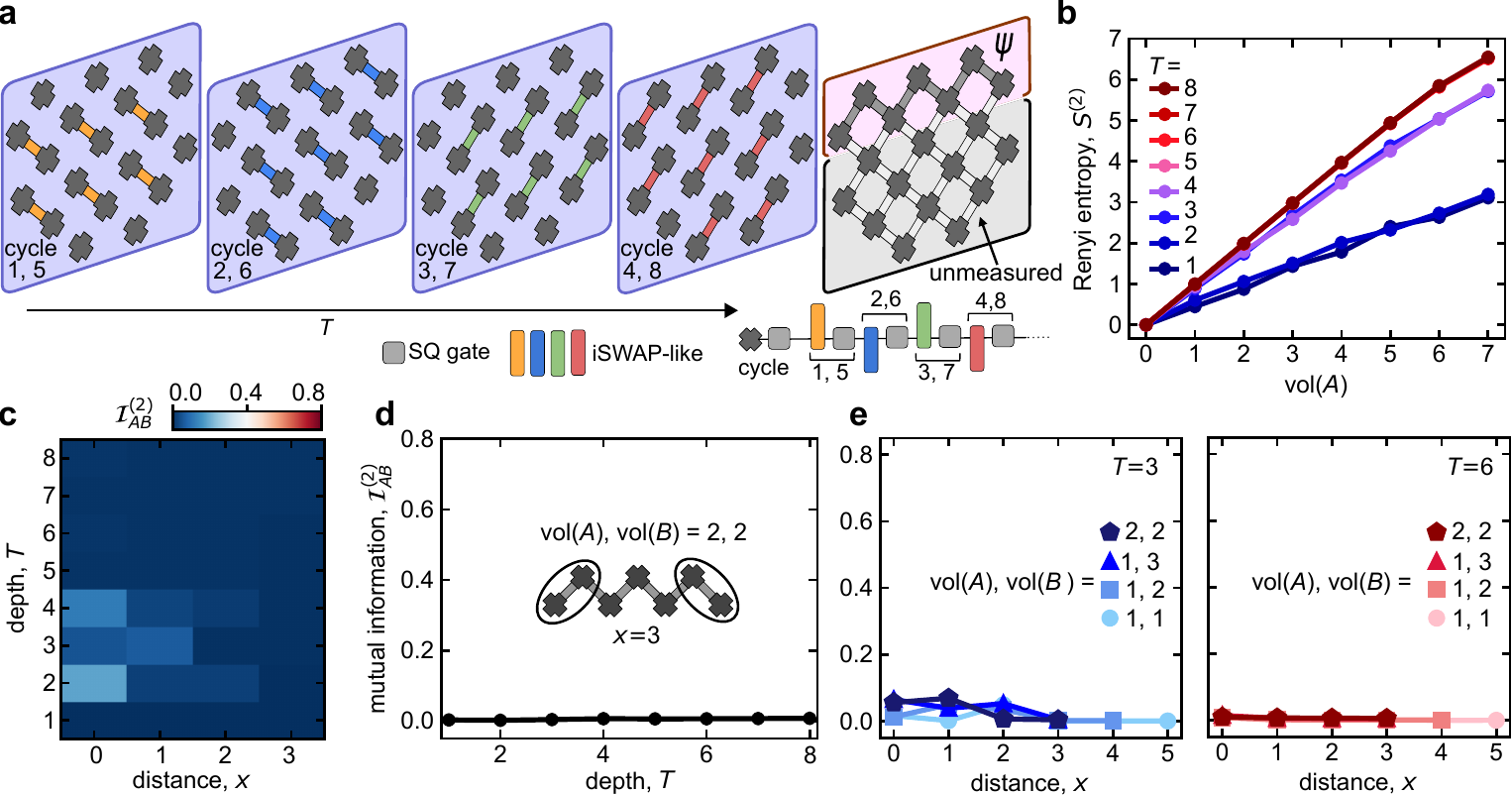} 
\caption{\small{\textbf{a}, Schematic of the 2D grid of qubits. At each cycle of the circuit, random single qubit and two-qubit iSWAP-like gates are applied to each qubit in the cycle sequence shown. At the end of the circuit, the measurement outcomes of the 12 lower $M$ qubits are \textit{not} postselected on. \textbf{b}, Second Renyi entropy of contiguous subsystems $A$ of the $L=7$ edge qubits at various depths. The measurement is not noise-mitigated here. \textbf{c}, Second Renyi mutual information $\mathcal{I}_{AB}^{(2)}$ between 2-qubit subsystems $A$ and $B$ vs depth $T$ and distance $x$ (number of qubits between $A$ and $B$). \textbf{d}, $\mathcal{I}_{AB}^{(2)}$ as a function of $T$ for 2-qubit subsystems $A$ and $B$ at maximum separation. \textbf{e}, \textbf{f}, $\mathcal{I}_{AB}^{(2)}$ vs. $x$ for $T = 3$ and $T = 6$, respectively, for different volumes of $A$ and $B$.}\label{fig:fig3_nps}}
\end{figure*}

We also present additional data complementary to Fig. 3 of the main text. In Fig.~\ref{fig:qmi_ab} we plot the second Renyi mutual information $\mathcal{I}_{AB}^{(2)}$ as a function of the total volume of the combined subsystems $A$ and $B$ of the 7 edge qubits.
When $T \geq 4$, $\mathcal{I}_{AB}^{(2)}$ is large if $A$ and $B$ together make up over half the system ($\geq 4$ qubits) and small otherwise ($\leq 3$ qubits), consistent with the expected behavior of highly entangled states.
Lastly, we again highlight the importance of postselection. In Fig.~\ref{fig:fig3_nps} we show the equivalent results to Fig. 3 of the main text, but without postselection. We find no signature of a transition between entanglement phases in the 1D chain of 7 qubits and correlations between qubits, as measured by the second Renyi mutual information, are entirely absent. 

\subsection{Theoretical justification of error mitigation prescription} \label{sec:mitgation_works}

Here we show that the prescription in Eq.~\eqref{eq:supp_error_mitigation_def} is theoretically justified for states prepared by short time evolution. 
Specifically, we show that, to leading order in the noise strength $\epsilon$, the entropy $S^{(2)}_A$ incurs an extensive error $\sim \epsilon N_A$ (where $N_A$ is the number of qubits in $A$), while the mitigated entropy incurs only a constant error, $\sim {\rm const.}\times \epsilon$. 

Let us for simplicity consider a quantum state $\ket{\psi}$ produced by the ideal, noiseless circuit, and affected by a local noise channel $\mathcal{E}$ only at the end of the dynamics (this approximation is reasonable as long as we focus on states prepared by short unitary evolution, where a large fraction of the noise comes from measurement error and even mid-circuit gate error may be commuted to the final time at the expense of finite-range correlations).
The randomized measurement protocol probes the entropy of the noisy state ${\rho}^{\rm noisy}_A = \Tr_{\bar A}[ \mathcal{E}^{\otimes N}(\ket{\psi}\bra{\psi})]$, while we aim to recover the entropy of the ideal state $\rho_A = \Tr_{\bar A} [\ketbra{\psi}{\psi}]$. Note that ${\rho}_A^{\rm noisy} = \mathcal{E}^{\otimes N_A} (\rho_A)$, where $N_A$ is the number of qubits in $A$, due to $\mathcal{E}$ being trace-preserving. 

We take the local noise channel $\mathcal{E}$ to be single-qubit depolarizing noise, which is often a good approximation in the context of random circuit dynamics~\cite{arute2019quantum}:
\begin{equation}
    \mathcal{E}_i (\rho_A ) = e^{\epsilon \mathcal{L}_i}(\rho_A), 
    \quad
    \mathcal{L}_i (\rho_A) = -3\rho_A + \sum_{\alpha = x,y,z}\sigma_i^\alpha \rho_A \sigma_i^\alpha = -4\rho_A + 2 I_i \otimes \rho_{A\setminus \{i\}} . \label{eq:supp_depolarizing}
\end{equation} 
Here $\epsilon$ parametrizes the noise strength and $\mathcal{L}_i$ is the Liouvillian generator of the noise channel, in the Gorini-Kossakowski-Sudarshan-Lindblad form with jump operators $\{\sigma_i^\alpha:\ \alpha = x,y,z\}$. We have additionally exploited the decomposition of the erasure channel $\frac{1}{2} I_i \otimes \rho_{A\setminus\{i\}}
= \frac{1}{4}\left(\rho_A + \sum_{\alpha = x,y,z} \sigma_i^\alpha \rho_A \sigma_i^\alpha \right)$ in the second equality. 

The entropy mitigation prescription Eq.~\eqref{eq:supp_error_mitigation_def} is to approximate the ideal entropy $S_A^{(2)} = -\log_2 \Tr(\rho_A^2)$ by the quantity 
\begin{equation}
    \tilde{S}_A^{(2)} = -\log_2 \Tr[({\rho}_A^{\rm noisy})^2] + \frac{N_A}{N} \log_2 \Tr[({\rho}^{\rm noisy})^2].
\end{equation}
To evaluate the effect of weak noise on the mitigated entropy, we differentiate with respect to $\epsilon$ at $\epsilon = 0$:
\begin{align}
    \left. \frac{d}{d\epsilon} \tilde{S}^{(2)}_A \right|_{\epsilon = 0}
    & = -\frac{2}{\ln(2)}\sum_{i = 1}^{N_A} \frac{\Tr[\rho_A \mathcal{L}_i (\rho_A)] }{\Tr \rho_A^2} 
    + \frac{2N_A}{\ln(2)N} \sum_{i=1}^N  \frac{\Tr[ \rho \mathcal{L}_i (\rho) ] }{\Tr \rho^2} \nonumber \\
    & = \frac{4N_A}{\ln(2)} \left\{ 
        \frac{1}{N_A} \sum_{i=1}^{N_A} \left[2-\frac{{\rm Tr} (\rho_{A\setminus\{i\}}^2 )}{\Tr\rho_A^2} \right]
        - \frac{1}{N} \sum_{i=1}^{N} \left[2- {\rm Tr} (\rho_{\{i\}}^2 ) \right] \right\} \nonumber \\
    & = -\frac{4N_A}{\ln(2)} \left\{ 
        \frac{1}{N_A} \sum_{i=1}^{N_A} \Tr(\rho_{\{i\}}^2) 2^{\mathcal{I}^{(2)}_{{\bar A}, \{i\}}}
        - \frac{1}{N} \sum_{i=1}^{N}  {\rm Tr} (\rho_{\{i\}}^2 )
        \label{eq:supp_dSdeps}
    \right\}
\end{align}
In the second line we have used the form of $\mathcal{L}_i$ from Eq.~\eqref{eq:supp_depolarizing} and the fact that $\rho = \ketbra{\psi}{\psi}$ is pure (note the expression is evaluated at $\epsilon = 0$) to simplify $\Tr \rho^2 = 1$ and $\Tr \rho_{ \overline{\{i\}} }^2 = \Tr \rho_{\{i\}}^2$.
In the third line we have used the definition of Renyi-2 mutual information, and again the fact that $\rho$ as a whole is pure, to simplify
\begin{equation}
    \frac{\Tr( \rho_{A\setminus \{i\}}^2) }{\Tr \rho_A^2}
    = \Tr( \rho_{\{i\}}^2) \left( \frac{\Tr( \rho_{\bar{A} \cup \{i\} }^2) }{\Tr \rho_{\{i\}}^2 \Tr \rho_{\bar A}^2 } \right)
    = \Tr( \rho_{\{i\}}^2) 2^{S^{(2)}_{\{i\}} + S^{(2)}_{\bar A} - S^{(2)}_{\bar A \cup \{i\}} }
    = \Tr( \rho_{\{i\}}^2) 2^{\mathcal{I}^{(2)}_{{\bar A}, \{i\}}}.
\end{equation} 
Finally, neglecting the position dependence of single-site purities and setting $\Tr\rho_{\{i\}}^2 \equiv f \in [1/2,1]$ for all $i$, we arrive at 
\begin{equation}
    \left. \frac{d}{d\epsilon} \tilde{S}^{(2)}_A \right|_{\epsilon = 0} 
    = \frac{4f}{\ln(2)} \sum_{i\in A} \left( 2^{\mathcal{I}^{(2)}_{\bar{A}, \{i\}}} - 1 \right) 
    \approx 4f \sum_{i\in A} \mathcal{I}^{(2)}_{\bar{A}, \{i\}}.
\end{equation}
The approximation holds to leading order in small mutual information.
Clearly for the case of unitary dynamics the mutual information is short-ranged, being bounded by the light cone. Even in the presence of measurements, a general stability condition requires the sum to remain finite in the limit of large $A$ in 1D systems~\cite{LiVijay_DPRE_2021}. 
Thus the error-mitigated entropy carries a {\it finite} error
\begin{equation}
    \left. \tilde{S}_A^{(2)}\right|_\epsilon - \left. S_A^{(2)}\right|_{\epsilon = 0} = C \epsilon + O(\epsilon^2)
\end{equation}
for small noise strength $\epsilon$, where $C$ is a constant. At the same time it is apparent from Eq.~\eqref{eq:supp_dSdeps} that the un-mitigated entropy carries an extensive error
\begin{equation}
    \left. S_A^{(2)}\right|_\epsilon - \left. S_A^{(2)}\right|_{\epsilon = 0} = C' N_A \epsilon + C''\epsilon + O(\epsilon^2)
\end{equation}
where $C'$, $C''$ are constants.
Thus, while the error mitigation prescription does not completely reproduce the ideal, noiseless behavior, in the case of states prepared by short time evolution (whether unitary or monitored) it does remove the dominant entropy density contribution coming from noise, and in particular allows the correct identification of area-law states.

\subsection{Limitations of error mitigation: highly-entangled states} \label{sec:mitigation_fails}

Here we complement the analysis above by illustrating an example of how the error mitigation may fail in highly-entangled states by producing significant distortions of the underlying noiseless behavior. 
As an extreme example, one may consider a Haar-random state $\ket{\psi}$ on a bipartite system $AB$ and compute the purity $\Tr[({\rho}_A^{\rm noisy})^2]$ after the state has been subjected to local depolarizing noise. Formally we may write this quantity as a matrix element of the noise super-operator:
\begin{equation}
    2^{-S_A^{(2)}} 
    = \Tr [({\rho}_A^{\rm noisy})^2]
    = (\chi_A e_B | \mathcal{E}^{\otimes 2N} |\rho^{\otimes 2})
\end{equation}
where $\rho = \ketbra{\psi}{\psi}$ is the pure Haar-random state, $\mathcal{E}$ is a single-qubit depolarizing channel as before, and $e,\chi$ are permutation operators acting on the two replicas of the Hilbert space: $e\ket{i}_1 \ket{j}_2 = \ket{i}_1 \ket{j}_2$ (identity permutation) and $\chi\ket{i}_1 \ket{j}_2 = \ket{j}_1 \ket{i}_2$ (transposition).
Furthermore we use the notation $|O)$ to represent an operator $O$ as a vector in a doubled Hilbert space, with inner product $(O|P) = \Tr(O^\dagger P)$.

The purity can be exactly averaged over Haar-random states by using the fact that 
\begin{equation} 
    \mathbb{E}_{\psi \sim {\rm Haar}}[\ketbra{\psi}{\psi}^{\otimes 2}] 
    = \frac{ e_A e_B + \chi_A \chi_B }{2^N (2^N +1)} .
\end{equation}
Next, we note that the noise super-operator $\mathcal{E}^{\otimes 2} $ leaves invariant both the ``super-ket'' $|e)$ (unitality of the depolarizing channel, $\mathcal{E}(I) = I$) and the ``super-bra'' $(e|$ (trace preservation, $\mathcal{E}^\ast(I) = I$). 
Using the inner product rules $(e|e) = (\chi|\chi) = d^2$ and $(e|\chi) = (\chi|e) = d$ where $d$ is the relevant Hilbert space dimension, this gives
\begin{align}
    \mathbb{E}_{\psi \sim {\rm Haar}} \left[ 2^{-S_A^{(2)}}  \right]
    & = \frac{(\chi_A e_B| \mathcal{E}^{\otimes 2N} |e_A e_B) + (\chi_A e_B| \mathcal{E}^{\otimes 2N} |\chi_A \chi_B)}{2^N (2^N +1)}  \nonumber \\
    & = \frac{2^{2N_B} (\chi_A|e_A) + 2^{N_B} (\chi_A|\mathcal{E}^{\otimes 2N_A} |\chi_A)} {2^N (2^N + 1)} \nonumber \\
    & = \frac{2^{N_B} + 2^{-N_A} (\chi_i|\mathcal{E}_i^{\otimes 2} |\chi_i)^{N_A}}{2^N + 1} .
\end{align}
In the last line we have used the fact that $\chi_A = \bigotimes_{i\in A} \chi_i$ to factor the matrix element $(\chi_A|\mathcal{E}^{\otimes 2N_A} |\chi_A)$ into on-site matrix elements.
It is now helpful to decompose $\chi_i = (I^{\otimes 2} + X^{\otimes 2} + Y^{\otimes 2} + Z^{\otimes 2})/2$ and note that $\mathcal{E}(I) = I$ while $\mathcal{E}(X) = e^{-\epsilon} X$, and the same for $Y$ and $Z$. Thus 
$(\chi_i| \mathcal{E}_i^{\otimes 2} |\chi_i) = 1 + 3e^{-2\epsilon}$, and we conclude 
\begin{align}
    \mathbb{E}_{\psi \sim {\rm Haar}} \left[ 2^{-S_A^{(2)}}  \right]
    & = \frac{2^{N_B} + [(1+3e^{-2\epsilon})/2]^{N_A}}{2^N + 1}.
\end{align}
Based on quantum typicality, the state-averaged result is also typical, i.e., the behavior of typical individual states from the Hilbert space should be well-described by the above formula, giving in the limit of large $N$
\begin{equation}
    2^{-S_A^{(2)}} \simeq 2^{-N_A} + \left( \frac{1+3e^{-2\epsilon}}{4} \right)^{N_A} 2^{-N_B}.
\end{equation}
While the ascending part of the Page curve (small $N_A$) is nearly unchanged, the descending part is significantly affected by noise, giving in particular and entropy $S^{(2)} \simeq N\log_2[4/(1+3e^{-2\epsilon})]$ for the whole system. 

Using the error mitigation prescription above, we subtract the corresponding entropy density to get
\begin{equation}
    2^{-\tilde{S}^{(2)}_A }
    \simeq 2^{-N_B} + \left(\frac{2}{1+3e^{-2\epsilon}}\right)^{N_A}.
\end{equation}
In the mitigated entropy $\tilde{S}^{(2)}_A$, the descending part of the Page curve (small $N_B$) is accurately reproduced, and in particular terminates at zero entropy (for $N_B = 0$) as is expected for a pure state. 
However, the ascending part of the Page curve (small $N_A$) has an incorrect slope, $S^{(2)}_A \simeq \sigma N_A$ with ``entropy density'' $\sigma = \log_2[(1+3 e^{-2\epsilon})/2] < 1$ for $\epsilon >0$. 
Moreover, the maximum of $S^{(2)}$ (which is at the half-system cut $N_A = N/2$ in the absence of noise) moves to $N_A \simeq N / (1+\sigma) \simeq N/2 + \epsilon (3/4\ln(2))N$ (the latter valid at small $\epsilon$). 
This asymmetry is qualitatively consistent with the observed behavior of our experimental error-mitigated data in Fig.~2 and 3 in the main text.

\newpage

\section{Space-time dual of fSim gates}\label{sec:stdual}

\begin{figure*}[t!]
    \centering
    \includegraphics[width=0.5\textwidth]{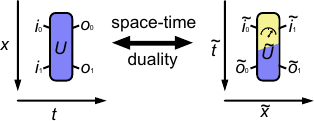} 
\caption{\small{Space-time dual transformation between a unitary gate $U$ and generically non-unitary gate $\tilde{U}$. The dimensions which act as ``space'' and ``time'' are swapped through this transformation: $x \leftrightarrow \tilde{t}, t \leftrightarrow \tilde{x}$.}\label{fig:spacetime_dual_gate}}
\end{figure*}

In Fig.~2 in the main text we use fSim gates with different parameters in order to realize different entanglement structures via space-time duality. Here we review the action of the space-time duality on fSim gates and obtain the types of measurements that appear in the effective monitored dynamics. 

If $i_o$ and $i_1$ label the qubit input states and $o_0, o_1$ the qubit output states, then the space-time duality transformation acts on a two-qubit unitary gate $U$ as a permutation of indices, namely
\begin{equation}
    U_{i_0,i_1}^{o_0, o_1} = \tilde{U}_{i_0, o_0}^{i_1, o_1}.
    \label{eq:std_def}
\end{equation}
In other words while $U$ maps input states $i_{0,1}$ to output states $o_{0,1}$, $\tilde{U}$ maps left states $i_0, o_0$ to right states $i_1, o_1$ (Fig.~\ref{fig:spacetime_dual_gate}). 
The linear map $\tilde{U}$ is generically non-unitary.

We focus on fSim gates with parameters $(\theta, \phi)$ that obey $\phi = 2\theta$, which interpolate between $\textsf{SWAP}$ ($\theta = \pi/2$) and identity ($\theta = 0$). The two gate sets used in the main text are $\theta = 2\pi/5$ and $\theta = \pi/10$. 
Using the definition in Eq.~\eqref{eq:std_def}, the space-time duality transformation acts on the matrix elements of the unitary gate $U$ as follows:
\begin{equation}
    U = 
    \begin{pmatrix}
        1 & 0 & {\color{blue} 0} & {\color{blue} 0} \\
        {\color{red} 0} & {\color{red} \cos\theta} & -i\sin\theta & 0 \\
        0 & -i\sin\theta & {\color{green} \cos\theta} & {\color{green} 0} \\
        {\color{orange} 0} & {\color{orange} 0} & 0 & e^{-2i \theta}
    \end{pmatrix}
    \mapsto 
    \tilde{U} = 
    \begin{pmatrix}
        1 & 0 & {\color{red} 0} & {\color{red} \cos\theta} \\
        {\color{blue} 0} & {\color{blue} 0}  & -i\sin\theta& 0 \\
        0 & -i\sin\theta & {\color{orange} 0} & {\color{orange} 0} \\
        {\color{green} \cos\theta} & {\color{green} 0} & 0 & e^{-2i\theta}
    \end{pmatrix}
    \label{eq:utilde_pmatrix}
\end{equation}
(the colors denote pairs of entries whose positions are being swapped under the transformation Eq.~\eqref{eq:std_def}: {\color{blue}blue} $\leftrightarrow$ {\color{red}red} and {\color{orange}orange} $\leftrightarrow$ {\color{green}green}).

One has $\tilde{U} = 2VH$ where $V$ is unitary and $H$ is Hermitian and satisfies $0\leq H^2 \leq I$~\cite{Ippoliti2021postselection}. This allows us to view $H$ as an outcome of a generalized measurement (namely an {\it instrument} in a positive-operator-valued measure, e.g. $\{H^2, I-H^2\}$).
From explicit calculation of $H = \sqrt{\tilde{U}^\dagger \tilde{U}}/2$ we obtain 
\begin{equation} 
H = \frac{1}{2}\sqrt{1+3\cos^2(\theta)} \ketbra{\psi_\theta} + \frac{1}{2} |\sin(\theta)| (I - \ketbra{\psi_\theta})
\end{equation}
with $\ket{\psi_\theta} \equiv \frac{1}{\sqrt 2} (e^{-i\theta}\ket{00} + e^{i\theta}\ket{11})$.

As consistency checks, one can see that $\theta = \pi/2$ yields $H = I/2$, i.e. there is no measurement: $\tilde{U} = V$. This is consistent with the fact that $\textsf{SWAP}$ is a self-dual gate.
On the opposite end, for $\theta = 0$ ($U = I$), one gets a projector $H = \ketbra{\psi_0}$ with $\ket{\psi_0} = (\ket{00} + \ket{11}) / \sqrt{2}$, a Bell pair state. 
Intermediate values of $\theta$ correspond to weak measurements of the operator $\ketbra{\psi_\theta}$.
The unitary $V$ is a composition of $\textsf{iSWAP}$ and a $2\times 2$ unitary operation $V'$ acting on the two-dimensional subspace spanned by $\ket{00}$ and $\ket{11}$. 
For the values of $\theta$ used in the main text, we have:
\begin{align}
    & H = 0.1545 I + 0.8090 \ketbra{\psi_{\pi/10}}, 
    & V' = \begin{pmatrix} 0.6487-0.3993i & 0.4935 + 0.4198i \\ 0.4935 + 0.4198i & 0.2901-0.7043i \end{pmatrix} \qquad (\theta = \pi/10), \\
    & H = 0.4755 I + 0.0916 \ketbra{\psi_{2\pi/5}},
    & V' =  \begin{pmatrix} 0.9585 - 0.0249i  & 0.2724 + 0.0807i \\ 0.2724 + 0.0807i & -0.7901 - 0.5432 i \end{pmatrix} \qquad (\theta = 2\pi/5).
\end{align}
The former value gives a stronger measurement ($H$ closer to a projector). 

\begin{figure}
    \centering
    \includegraphics[width = 0.5\textwidth]{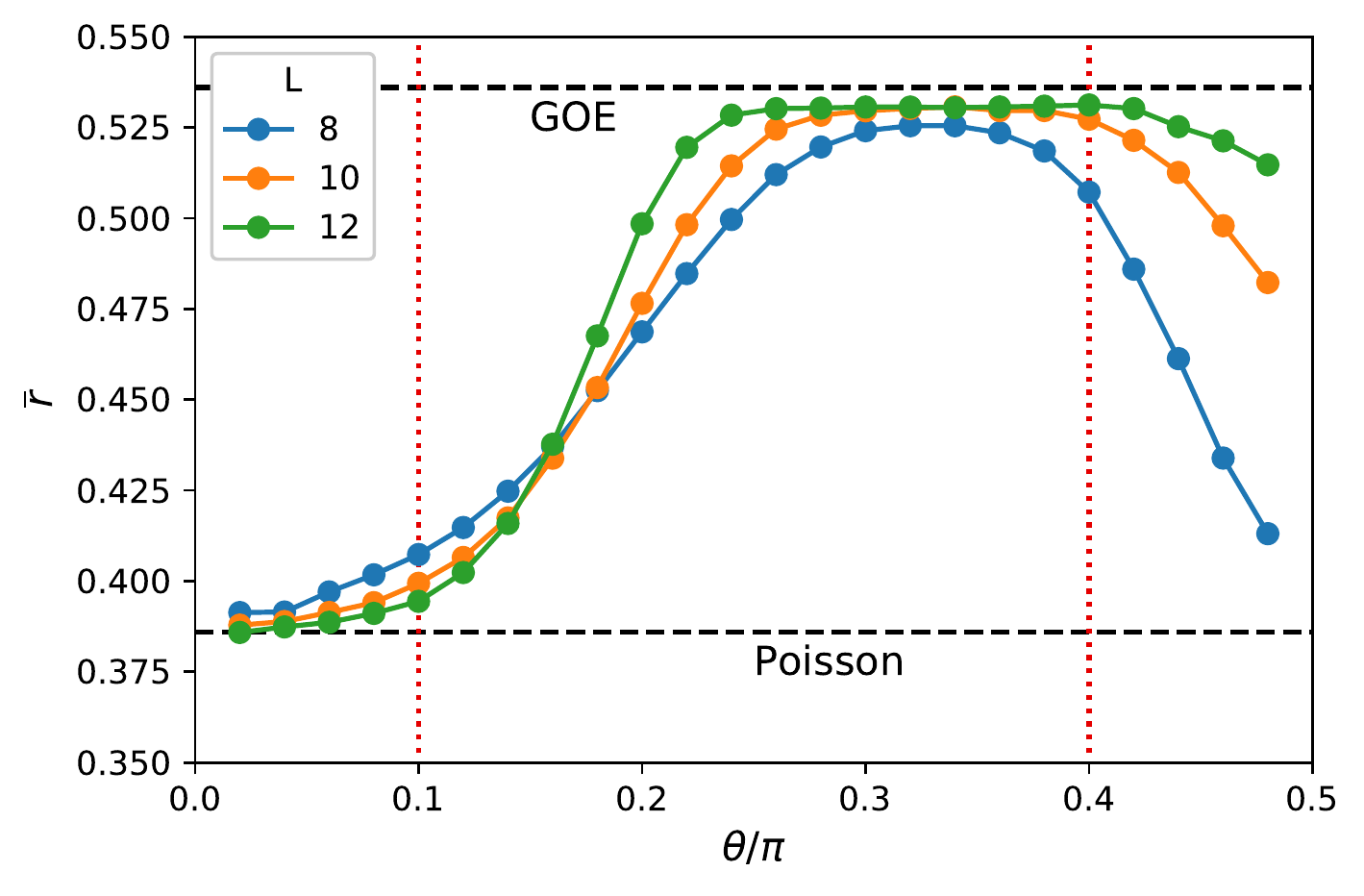}
    \caption{Mean level spacing ratio $\bar{r}$ for the quasienergy spectrum of the Floquet operator $U_F(\theta)$, Eq.~\eqref{eq:UFloquet}, in the half-filling charge sector, as a function of fSim angle $\theta$. Data averaged over at least $10^3$ realization of disorder for each system size. Vertical dotted lines represent the values used in the main text, $\theta = 0.1\pi$ and $\theta = 0.4\pi$. Horizontal dashed lines represent the reference values for Poisson (localized) and GOE (ergodic) models.}
    \label{fig:level_spacing}
\end{figure}

Finally, we verify that the two values of $\theta$ used in the main text giver rise to localized and ergodic unitary dynamics, respectively.
We consider time-periodic dynamics generated by the Floquet unitary 
\begin{equation}
    U_F(\theta) = e^{-i \sum_i h_i Z_i} \bigotimes_{i \text{ even}} \textrm{fSim}_{i,i+1}(\theta,2\theta) e^{-i \sum_i h_i Z_i} \bigotimes_{i \text{ odd}} \textrm{fSim}_{i,i+1} (\theta,2\theta) 
    \label{eq:UFloquet}
\end{equation}
on a 1D chain of $L$ qubits with open boundary conditions.
The $h_i$ are random fields drawn uniformly from $[0,2\pi)$. 
As the dynamics possesses a $U(1)$ symmetry (generated by $\sum_i Z_i$), the Hilbert space breaks up into decoupled charge sectors. We focus on the largest such sector, $\sum_i Z_i = 0$ (we take $L$ even). 
We then diagonalize $U_F(\theta)$ restricted to this sector, obtaining eigenvalues $\{e^{-iE_n}\}$ and thus a set of {\it quasi-energies} $\{E_n\}$. 
A standard diagnostic of ergodicity is the mean level spacing ratio $\bar{r}$ defined as the average over $n$ of
\begin{equation}
    r_n = \frac{\min(\delta_n, \delta_{n+1})}{\max(\delta_n, \delta_{n+1})}, 
    \quad 
    \delta_n \equiv E_{n+1} - E_n.
\end{equation}
We assumed the $E_n$ are sorted in ascending order. 
The quantity $\bar{r}$ takes on universal values $\bar{r} = 0.386\dots$ in the localized phase (Poisson level statistics) and $\bar{r} = 0.536\dots$ in the ergodic phase with time reversal symmetry (Gaussian Orthogonal Ensemble (GOE) level statistics).
Fig.~\ref{fig:level_spacing} shows numerical results for $\bar{r}$ as a function of $\theta$ for chains of $L = 8,10,12$ sites, showing a crossover between a localized and an ergodic regime at $\theta \simeq 0.15\pi$ at these sizes. 
Therefore, for the system size used in our experiment ($L = 7$, Fig.2 in the main text), the values $\theta = \pi/10$ and $\theta = 2\pi/5$ are firmly in the localized and ergodic regimes, respectively.

\newpage

\section{Mapping of $2+1$-dimensional shallow circuits to $1+1$-dimensional monitored circuits}\label{sec:mapping}

Here we provide details on the mapping between $(2+1)$-dimensional shallow circuits with final measurements to $(1+1)$-dimensional dynamics with mid-circuit measurements, following Ref.~\cite{Napp2022}.

\begin{figure}
    \centering 
    \includegraphics[width = \textwidth]{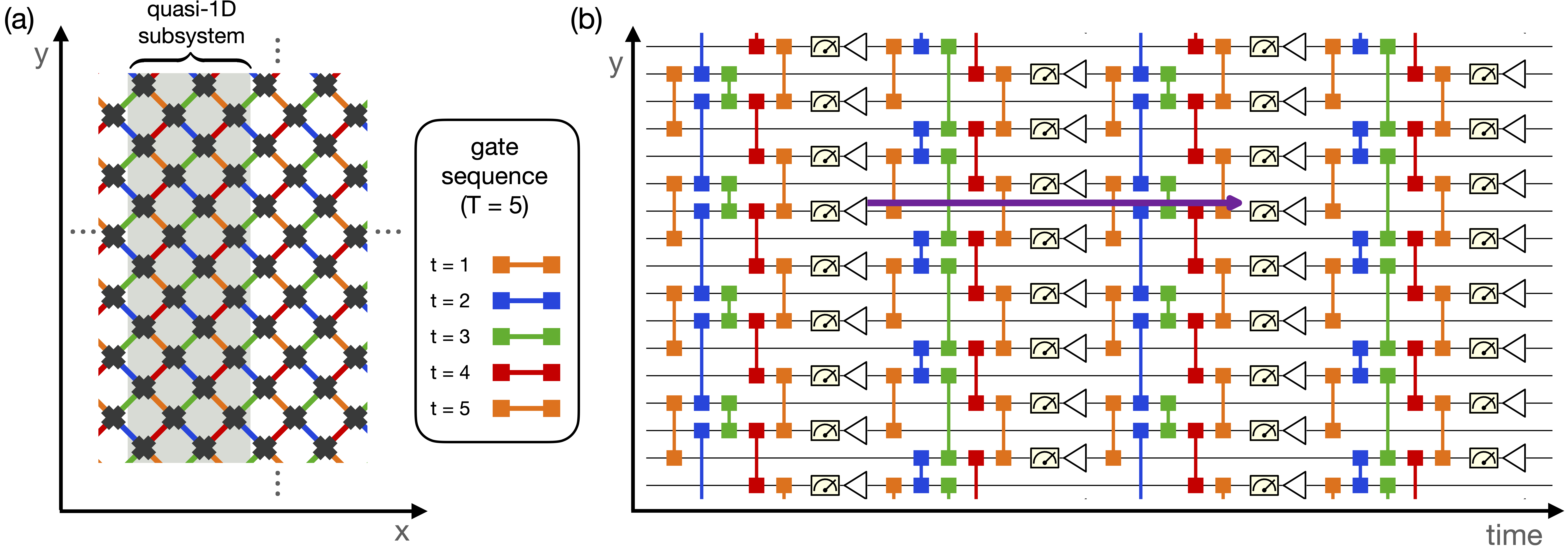}
    \caption{Mapping of $(2+1)$-dimensional shallow circuits to $(1+1)$-dimensional monitored circuits. (a) 2D circuit architecture and gate sequence. (b) Equivalent 1D monitored circuit. Each measurement is followed by a reset to the $\ket{0}$ state (white triangles). The purple arrow represents the ``life cycle'' of a qubit in the dynamics, from initialization in $\ket{0}$ through the depth-5 gate sequence ({\color{orange} orange}, {\color{blue} blue}, {\color{green} green}, {\color{red} red}, {\color{orange} orange}), to final measurement. The circuit architecture is periodic in space and time.} \label{fig:mapping}
\end{figure}

We consider the circuit architecture of Fig.~\ref{fig:mapping}(a), which is spatially-infinite and has depth $T = 5$. 
We take a quasi-1D system oriented along the $y$ direction, composed of two consecutive columns in the 2D system (Fig.~\ref{fig:mapping}(a)). The $(2+1)$D circuit may be exactly implemented by a sequence of gates, measurements and resets on this quasi-1D system. With each step, the quasi-1D subsystem moves in the $x$ direction, eventually ``sweeping'' the whole 2D lattice (this mapping is the basis of the ``space-evolving block decimation'' algorithm for sampling shallow circuits of Ref.~\cite{Napp2022}).
The resulting $(1+1)$D circuit for our case is shown in Fig.~\ref{fig:mapping}(b). 
The spatial range of the two-qubit gates is proportional to $T$; for $T=5$ we find up to third-nearest-neighbor gates. This shows that the mapping is only useful for shallow circuits (deep circuits would give rise to non-local interactions).

\begin{figure}
    \centering
    \includegraphics[width = 0.6\textwidth]{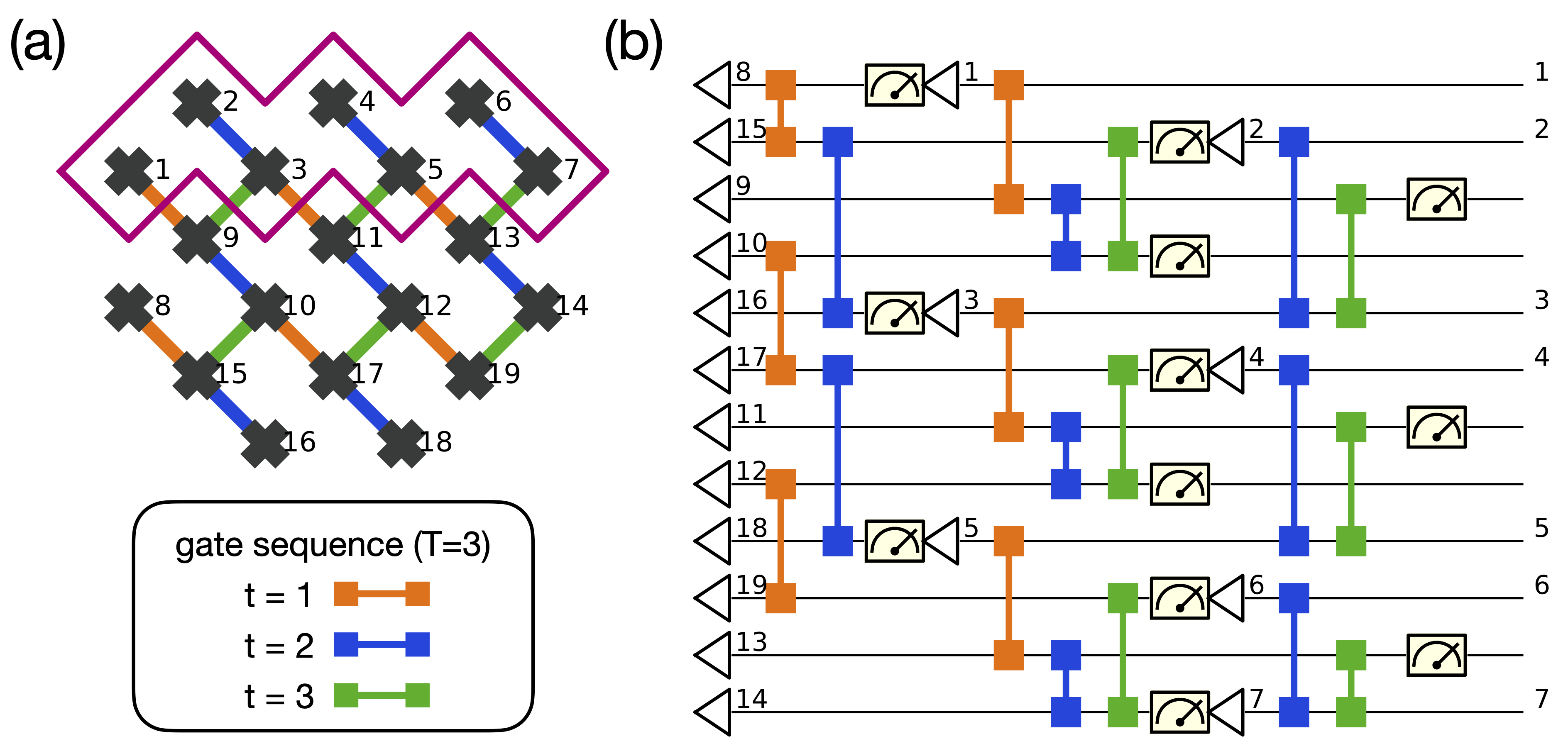}
    \caption{Exact mapping of 2D shallow circuit used in Fig.~3 of the main text, for depth $T = 3$, to 1D monitored circuit. 
    (a) Shallow 2D circuit architecture. Qubits are numbered from 1 to 19; qubits $Q_1$ through $Q_7$ (purple contour) make up the final state.
    (b) Equivalent 1D monitored circuit. White triangles represent resets to $\ket{0}$ states after measurement. Numbers on each qubit wire indicate the corresponding location in the 2D circuit (panel a). The circuit acts on 12 qubits but the final state involves only 7 of them (indicated by numbers on the right), the remaining 5 are decoupled by measurements.}
    \label{fig:mapping_19q}
\end{figure}

In addition, we show in Fig.~\ref{fig:mapping_19q} a specific instance of this mapping for the 19-qubit subsystem used in Fig.~3 of the main text. 

\newpage

\section{Decoding protocol}\label{sec:decoding}

\begin{figure*}[t!]
    \centering
    \includegraphics[width=0.97\textwidth]{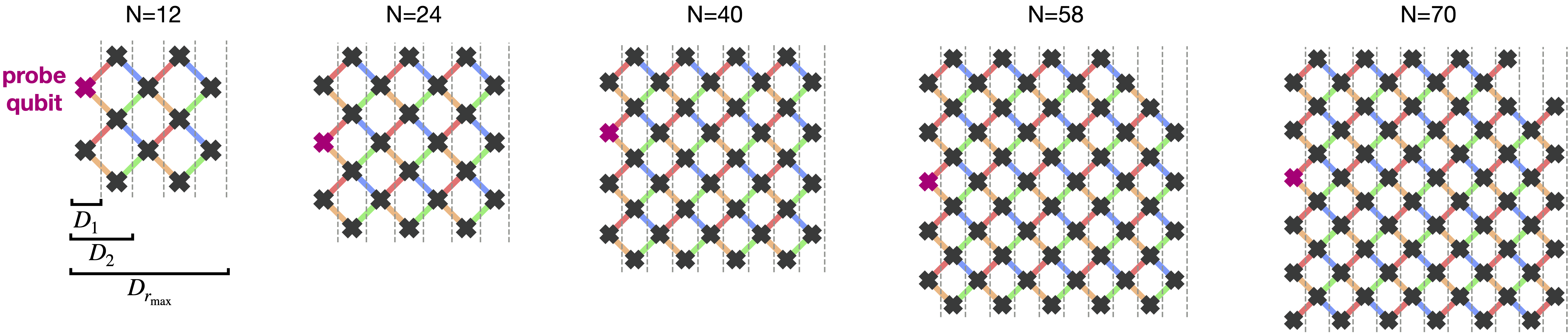} 
\caption{\small{The approximately square geometries of $N = 12, 24, 40, 58,$ and $70$ qubits used in the decoding and finite size analysis experiment of Fig. 4 in the main text. The bond colors denote the gate sequence (({\color{orange} orange}, {\color{blue} blue}, {\color{green} green}, {\color{red} red}, {\color{orange} orange}).
For each grid we show the probe qubit and the decoding patches $D_1$, $\dots D_{r_{\rm max}}$ employed in the decoding algorithm (dashed lines). \label{fig:setup}}}
\end{figure*}

The data in Figure 4 (main text) is obtained by decoding the entanglement of a local probe qubit~\cite{GullansPRL2020} on a 2D grid of $N = 12, 24, 40, 58,$ and $70$ qubits (Fig~\ref{fig:setup}). Here we briefly review the general idea of decodable order parameters for measurement-induced phases, the details of our specific protocol, and the classical simulation method used to implement it.

\subsection{Review}
As discussed in Ref.~\cite{GullansPRL2020}, the entanglement phases may be diagnosed in a scalable way by means of a probe qubit. In the original protocol, the probe qubit is prepared in a maximally-entangled state with the system a time $t = 0$; as time progresses, the monitoring in the system enventually disentangles the probe. However, in the disentangling (area-law) phase, this process takes a short, $N$-independent amount of time, while in the entangling (volume-law) phase it takes a very long time, diverging exponentially in $N$. Thus the probe qubit's entanglement at a time $t\sim N$ serves as an order parameter for the transition.

In a given quantum trajectory of the monitored evolution, labelled by the measurement record $m$, the probe qubit is in a state 
$ \rho_m = (I + \vec{a}_m \cdot \vec{\sigma})/2 $.
This state is fully described by the Bloch vector, $\vec{a}_m = \Tr(\vec{\sigma}\rho_m)$. The probe qubit's entanglement in trajectory $m$ is given by, e.g., its second Renyi entropy: $S^{(2)} = -\log_2[(1+a_m^2)/2]$ ($a_m = |\vec{a}_m|$ is the length of the vector). 
Thus the average length of the Bloch vector, $\mathbb{E}_m[a_m]$, can be used to distinguish the two phases.

Trajectory-averaged observables provide no information about this: indeed, $\mathbb{E}_m \Tr(\rho_m \vec{\sigma}) = \mathbb{E}_m[\vec{a}_m] = \vec{0}$ due to the random direction of the vectors $\vec{a}_m$, independently of their lengths.
To access the length of the vectors experimentally, Ref.~\cite{GullansPRL2020} proposes using a conditional rotation of the probe, $U_m$, that rotates its Bloch vector along the $+\hat{z}$ direction: $U_m \rho_m U_m^\dagger = (I + a_m \hat{z})/2$. Following this conditional operation, the probe is measured in the computational basis yielding
$\mathbb{E}_m [\Tr(U_m\rho_mU_m^\dagger \sigma^z)]= \mathbb{E}_m[a_m]$. 
Such active feedback would require a classical computation of the conditional operation $U_m$ in real time, which is challenging on present-day quantum simulators due to the short coherence time. To avoid this issue, Ref.~\cite{NoelNatPhys2021} uses specially-structured Clifford circuits with pre-compiled feedback operations. 
We instead develop an approach that works in more general (non-Clifford) circuits, in which the ``feedback'' step is implemented virtually during a classical post-processing step.
In essence, we apply a classical bit flip that on average aligns the $z$ component of $\vec{a}_m$ along $+\hat{z}$; this allows us to extract the average $\mathbb{E}_m[|a^z_m|]$ of the component. Then, using the fact that different instances of the random circuit randomize the direction of $\vec{a}_m$, we connect this to the desired quantity $\mathbb{E}_m[a_m]$, as detailed below.

\subsection{Protocol}

Each iteration $i$ of the experiment produces a bitstring $\mathbf{z}^{(i)} = (z_1^{(i)}, \dots z_N^{(i)})$, with $z_n^{(i)} = \pm 1$ for each qubit in the system (note here we use binary values $\pm 1$ instead of $0,1$ for convenience).
We pick a qubit in the system to act as a ``probe'', $P$, and a set of qubits around it to be the ``decoding patch'', $D$. Remaining qubits are traced out and not used in the decoding protocol, serving as an ``environment'' that the probe may get entangled with. 
We emphasize that the probe qubit $P$ is on the same footing as all other qubits, and can be freely chosen after the experiment, during classical post-processing.

In each iteration of the experiment, we refer to the bits $z_n^{(i)}$ from qubits in the decoding patch $D$ as the {\it measurement record} $m^{(i)}$, while the bit on the probe qubit is $z_P^{(i)}$. 
Given the circuit architecture, gates, and a measurement record $m$, it is possible to classically compute the state $\rho_m^{\rm sim}$ and thus the Bloch vector $\vec{a}^{\,\rm sim}_m$. Note this step may be computationally hard in general; see Sec.~\ref{sec:supp_classicalsim} for details on the classical simulation.
From each shot of the experiment we define a classical bit, $\tau_m = \textsf{sign}(\vec{a}^{\,\rm sim}_m \cdot \hat{z})$ (i.e., $+1$ if $\vec{a}^{\,\rm sim}_m$ points in the northern hemisphere of the Bloch sphere, $-1$ otherwise). 
Finally, we compute the quantity
\begin{equation}
    \zeta^{(K)} = \frac{2}{K} \sum_{i=1}^K z_P^{(i)} \tau_{m^{(i)}}
\end{equation}
where $i = 1, \dots K$ runs over iterations of the experiment.
The expected value of $\zeta$ after averaging over many realizations of the experiment is
\begin{align}
    \zeta^{(K)} \xrightarrow{K\to\infty} \zeta 
    & = 2 \sum_m  \sum_{z_P = \pm 1} \textsf{Prob}(z_P,m)\, \textsf{sign}(a_m^{z, {\rm sim}}) z_P  \nonumber \\
    & = 2 \sum_m \textsf{Prob}(m)\, \textsf{sign}(a_m^{z, {\rm sim}}) \sum_{z_P = \pm 1} {\sf Prob}(z_P|m)  z_P  \nonumber \\
    & = 2 \sum_m \textsf{Prob}(m)\, \textsf{sign}(a_m^{z, {\rm sim}}) \sum_{z_P = \pm 1} \frac{1+z_P a_m^z}{2}\,  z_P \nonumber \\
    & = 2 \sum_m \textsf{Prob}(m)\, \textsf{sign}(a_m^{z, {\rm sim}})\, a_m^z.
    \label{eq:supp_zeta_def}
\end{align}
Here $a_m^{z} = \vec{a}_m \cdot \hat{z}$ denotes the $z$ component of the vector, we have introduced the conditional probability $\textsf{Prob}(z_P|m) = \textsf{Prob}(z_p,m) / \textsf{Prob}(m)$, and used the fact that $\textsf{Prob}(z_P = \pm 1|m) = \frac{1\pm a_m^z}{2}$.
Now, assuming the classical simulation is sufficiently accurate that we can substitute $\textsf{sign}(a_m^{z, {\rm sim}}) \mapsto \textsf{sign}(a_m^z)$, this reduces to 
\begin{equation}
    \zeta = 2\sum_m \textsf{Prob}(m) \overbrace{a_m^z \textsf{sign}(a_m^z)}^{|a_m^z|} = 2 \mathbb{E}[|a_m^z|]. \label{eq:supp_zeta_am}
\end{equation}
The average of the $z$ component $|a_m^z|$ is correlated with the average length $a_m$. This relationship can be sharpened by averaging over random circuit realizations, and assuming the direction of $\vec{a}_m$ on the Bloch sphere is uniformly random (this holds if the circuit contains a final layer of random unitary rotations). 
For a unit vector $\hat{n}$ distributed uniformly on the unit sphere, one has
\begin{equation}
    \mathbb{E}[|n^z|] 
    = \frac{1}{2} \int_{-\pi/2}^{\pi/2} {\rm d}\theta \sin\theta |\cos\theta|
    = \frac{1}{2}.
\end{equation}
Thus, averaging over circuit realizations $U$ as well as measurement records $m$, we get $\mathbb{E} [|a_m^z|] = \frac{1}{2} \mathbb{E}[a_m]$, and in conclusion
\begin{equation}
    \zeta = \mathbb{E}[a_m].
\end{equation}

We can then define a proxy for the probe's entropy, $S_{\rm proxy}^{(2)} = -\log_2[(1+\zeta^2)/2] = -\log_2[(1+(\mathbb{E}|a_m|)^2)/2]$. 
Note that this differs from both the ``quenched'' average of the second Renyi entropy, $S^{(2)} = - \mathbb{E} \log_2[(1+a_m^2)/2]$ and the ``annealed'' average $S^{(2)} = -\log_2[(1+\mathbb{E}[a_m^2])/2]$. 
It is true nonetheless that $S^{(2)}_{\rm proxy} = 0$ (1) if and only if the probe qubit is disentangled (maximally entangled) with unit probability.

We decode the same experimental dataset in multiple ways, corresponding to different decoding patches $D_r$.
These are indexed by a {\it decoding radius} $r = 0, 1, \dots r_{\rm max}$ such that $D_0$ is empty, $D_{r_{\rm max}}$ is the whole system minus the probe, and for each $r$ we have $D_r \subset D_{r+1}$. The patches used for the various system sizes are illustrade in Fig.~\ref{fig:setup}.
This defines $r$-dependent quantities $\zeta(r)$ and $S_{\rm proxy}^{(2)}(r)$.
The decoding radius $r$ in our shallow 2D circuits plays a role analogous to time in $1+1$-dimensional monitored circuits: persistence of the probe's entanglement up to large $r$ diagnoses the entangling phase.
In addition, it has a natural interpretation in terms of teleportation: by measuring and decoding the outcomes, entanglement may be generated between the probe qubit and the un-measured qubits a distance $r$ away, even if $r$ exceeds the size of the unitary light cone~\cite{Altman2021Teleportation}. 
This point of view further highlights the need to make use of the measurement record, since the measurement-averaged dynamics is strictly causal and cannot give rise to teleportation.

\subsection{Classical simulation algorithm}\label{sec:supp_classicalsim}

All the data presented in the main text is obtained from exact simulations of the relevant circuits. While this is expected to become impractical as the circuits approach a beyond-classical regime~\cite{arute2019quantum}, it is viable for the circuit sizes studied here ($N\leq 70$ qubits, depth $T=5$); see Sec.~\ref{sec:approx_decoding} for a scalable approach that can overcome this this limitation in one of the phases.

Rather than simulating an $N$-qubit wavefunction (which becomes prohibitive well below $N=70$), we simulate ``slices'' of the circuit by starting from the probe qubit and moving in the direction of increasing $r$, injecting new qubits and measuring them out as we go, similar to the approach of Ref.~\cite{Napp2022}. This allows us to obtain the desired quantity $\tau_m = \pm 1$ for an $N$-qubit experiment without storing in memory a whole $N$-qubit wavefunction. The classical computational cost is asymptotically $\exp(O(T\sqrt{N}))$, much lower than $\exp(O(N))$ as long as the circuit is shallow (small $T$). 
In practice, the precise cost depends sensitively on the choice of probe qubit and decoding patches; in our case, for $N=70$ with the decoding patches of Fig.~\ref{fig:setup}, it is possible to directly simulate the 1D circuit in Fig.~\ref{fig:mapping} on 24 qubits. However, further optimization of the gate order reduces this to 20 qubits, and presents an algorithm applicable to general geometries and decoding patches, described in detail below.

The algorithm is based on a wavefunction $\ket{\psi}_A$ defined on an ``active subsystem'' $A$ which always contains the probe $P$, but otherwise changes over the course of the algorithm: new qubits are introduced in the initial state $\ket{0}$ while other qubits are measured out, i.e. projected on basis states $\bra{m_j}$ determined by the experimental measurement record $m$ for the given shot. As it sweeps through the quantum processor, the algorithm sequentially obtains the bits $\tau_m (r)$, with $r$ increasing from 0 to $r_{\rm max}$; these are then used in conjunction with the experimental measurements $z_P$ to get the $r$-dependent order parameter $\zeta(r)$. We use $S$ to denote the entire system, i.e. the collection of all qubits in the processor: $S = \{Q_1, \dots Q_N\}$.

\begin{enumerate}
    \item {\bf Setup}: Choose a probe qubit $P \in S$ and decoding patches $\{D_r\}_{r=0}^{r_{\rm max}}$.
    \item {\bf Initialize}: set the {\it active subsystem} to $A = \{P\}$, the {\it initial-state subsystem} to $I = S\setminus P$, the {\it final-state subsystem} to $F = \emptyset$, the {\it wavefunction} to $\ket{\psi}_A = \ket{0}_P$
    \item {\bf Step $r = 0$}: 
    \begin{enumerate}[label = \Alph*.]
        \item \underline{Grow}: 
        \begin{enumerate}[label = (\roman*)]
            \item obtain the past lightcone\hyperlink{bldef}{*} of the probe, $L(P)$
            \item add all qubits in $L(P)$ to the active subsystem: $A \leftarrow A \cup L(P)$
            \item remove them from the initial-state subsystem: $I \leftarrow I \setminus L(P)$
            \item update the wavefunction: $\ket{\psi}_A \leftarrow \ket{\psi}_A \otimes \ket{0}_{L(P)}$
        \end{enumerate}
        \item \underline{Evolve}: apply all allowed gates\hyperlink{agdef}{${}^\ddagger$} to $\ket{\psi}_A$.
        \item \underline{Collect data}: evaluate and store $\tau_m (0) = \textsf{sign}(\bra{\psi}\sigma_P^z \ket{\psi})$
        \item \underline{Move to next step}: $r \leftarrow r+1$
    \end{enumerate}
    \item {\bf Steps} $0<r\leq r_{\rm max}$:
    \begin{enumerate}[label = \Alph*.]
        \item \hypertarget{step4a}{\underline{Grow}}: 
        \begin{enumerate}[label = (\roman*)]
            \item obtain the past lightcone\hyperlink{bldef}{*} $L(Q_n)$ of each qubit $Q_n \in D_r \setminus F$
            \item \hypertarget{greedy}{choose} qubit $Q_{n^\ast}$ with $n^\ast = {\rm argmin}_n |L(Q_n) \cap I|$ (i.e. the qubit whose past lightcone\hyperlink{bldef}{*} includes the fewest initial-state qubits)
            \item add all initial-state qubits in $L(Q_{n^\ast})$ to the active subsystem: $A \leftarrow A \cup (L(Q_{n^\ast}) \cap I)$
            \item update the wavefunction: $\ket{\psi}_A \leftarrow \ket{\psi}_A \otimes \ket{0}_{L(Q_{n^\ast})\cap I}$
            \item remove the qubits from the initial-state subsystem: $I \leftarrow I \setminus L(Q_{n^\ast})$
        \end{enumerate}
        \item \underline{Evolve}: apply all allowed gates\hyperlink{agdef}{${}^\ddagger$} to $\ket{\psi}_A$.
        \item \underline{Project}: for any qubit $Q_j \in A \cap D_r$ with no gates left to apply, 
        \begin{enumerate}[label = (\roman*)]
            \item remove it from the active subsystem,  $A \leftarrow A \setminus \{Q_j\}$
            \item add it to the final-state subsystem, $F \leftarrow F \cup \{Q_j\}$
            \item update the wavefunction by projecting according to the experimental measurement outcome $m_j$: $\ket{\psi}_A \leftarrow \langle m_j | \psi\rangle_A / \| \langle m_j| \psi\rangle_A \|$
        \end{enumerate}
        \item[D.] If $D_r \subseteq F$ (all qubits in the decoding patch $D_r$ have been measured out),
        \begin{enumerate}[label = (\roman*)]
            \item \underline{Collect data}: evaluate and store $\tau_m (r) = \textsf{sign}(\bra{\psi} \sigma_P^z \ket{\psi})$
            \item \underline{Move to next step}: $r \leftarrow r+1$
        \end{enumerate}
        \item[D'.] Else: go to step \hyperlink{step4a}{4A}.
    \end{enumerate}
    \item {\bf End} when $r > r_{\rm max}$. Return values of $\{\tau_m (r)\}_{r=0}^{r_{\rm max}}$.
\end{enumerate}
Definitions used in the algorithm:
\begin{enumerate}
\item[*] \hypertarget{bldef}{Past lightcone}: the past lightcone $L(Q_n)$ of a qubit $Q_n$ is the subset of qubits that can be reached starting from $Q_n$ at the final time $t = T$ and hopping along gates in the circuit for times $t = T, T-1, \dots 1$. It is the minimal subsystem needed to simulate the state (reduced density matrix) of qubit $Q_n$ at the end of the circuit.
\item[$\ddagger$] \hypertarget{agdef}{Allowed gates}: a gate acting on qubits $Q_i$, $Q_j$ in the first layer of the circuit ($t = 1$) is {\it allowed} if both qubits $Q_{i,j}$ belong to the active subsystem $A$. Gates in following layers $t = 2,\dots T$ are allowed if the same criterion is met {\it and} all gates on $Q_{i,j}$ in layers $\tau < t$ have already been applied.
\end{enumerate}

\begin{figure}
    \centering
    \includegraphics[width=\linewidth]{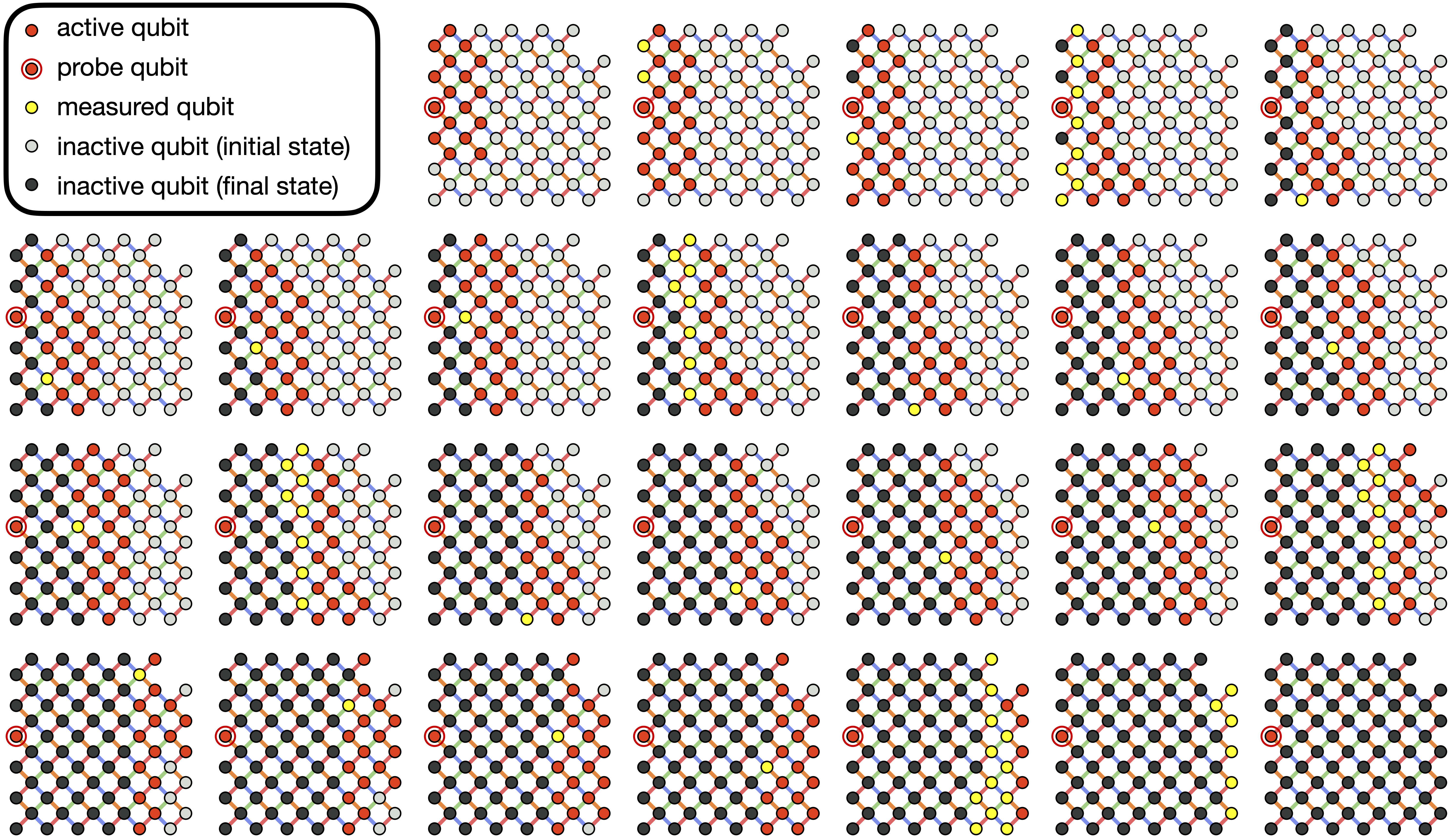}
    \caption{Classical simulation algorithm for quantum circuits on $N=70$ qubits (depth $T = 5$).
    The active subsystem $A$ is indicated in red. It always includes the probe $P$ (circled qubit). Light gray qubits make up the initial-state subsystem $I$ (not yet part of the wavefunction), black qubits are the final-state subsystem $F$ (already projected out and not part of the wavefunction anymore). Yellow qubits are projected out right before the evolution step in each panel. The active subsystem sweeps the chip from left to right and the simulation terminates when all qubits (except the probe) have been measured out. At each step in the simulation, the active subsystem contains at most 20 qubits. \label{fig:supp_N70_algo}}
\end{figure}

The algorithm splits the system $S$ into subsystems $I$ (initial), $A$ (active) and $F$ (final). All qubits (except the probe) start in $I$, move to $A$ and end in $F$. When $F$ coincides with a decoding patch $D_r$, we can evaluate the classical bit $\tau_m (r)$ needed for decoding. 
A concrete instance of the algorithm is shown in Fig.~\ref{fig:supp_N70_algo} for the largest circuit decoded in this work ($N=70$ qubits, depth $T = 5$, decoding patches $\{D_r:\, 0\leq r\leq r_{\rm max}=12\}$ as in Fig.~\ref{fig:setup}).
The color coding shows which qubits are measured out before each step, in addition to the three subsystems. Note that in several cases, after a patch $D_r$ is complete, we may immediately measure out some of the qubits in $D_{r+1}$; in these cases, Fig.~\ref{fig:supp_N70_algo} combines the two steps in a single panel.

Finally, we note that step \hyperlink{greedy}{4.A(ii)} in the algorithm corresponds to a greedy optimization of the tensor contraction order. While this works well in practice for the circuits studied here, in principle one may use better optimization strategies to push this decoding approach close to the beyond-classical frontier.

\subsection{Simulated decoding data}

\begin{figure}
    \centering
    \includegraphics[width=0.8\textwidth]{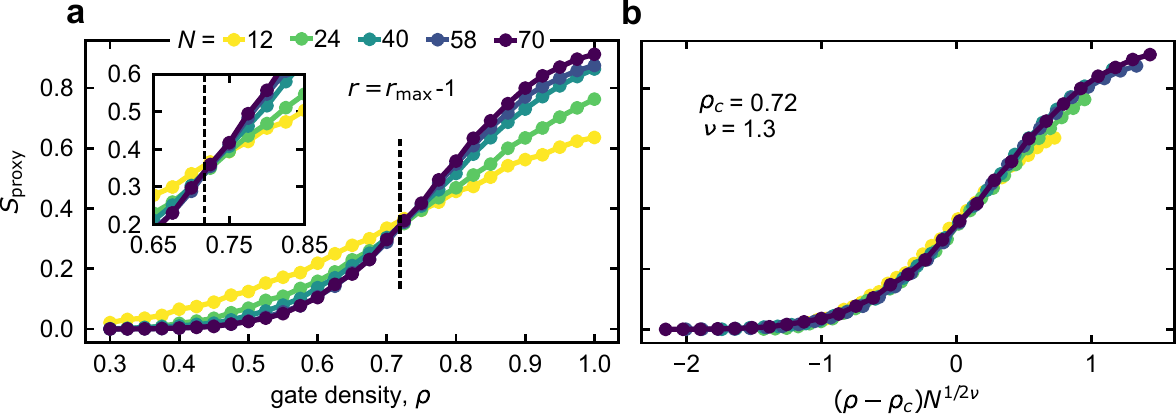}
    \caption{Simulated decoding of probe qubit (cf. Fig.~4 in the main text). 
    (a) Simulated proxy entropy $S_{\rm proxy}$ at decoding radius $r = r_{\rm max}-1$ as a function of $\rho$ for all system sizes studied. Vertical dashed line indicates the estimated $\rho_c = 0.72$. Inset: zoomed-in finite-size crossing.
    (b) Scaling collapse of the data vs $(\rho - \rho_c) N^{1/(2\nu)}$ with $\rho_c = 0.72$ and $\nu = 1.3$. 
    }
    \label{fig:simdecoding}
\end{figure}

\begin{figure*}[t!]
    \centering
    \includegraphics[width=0.97\textwidth]{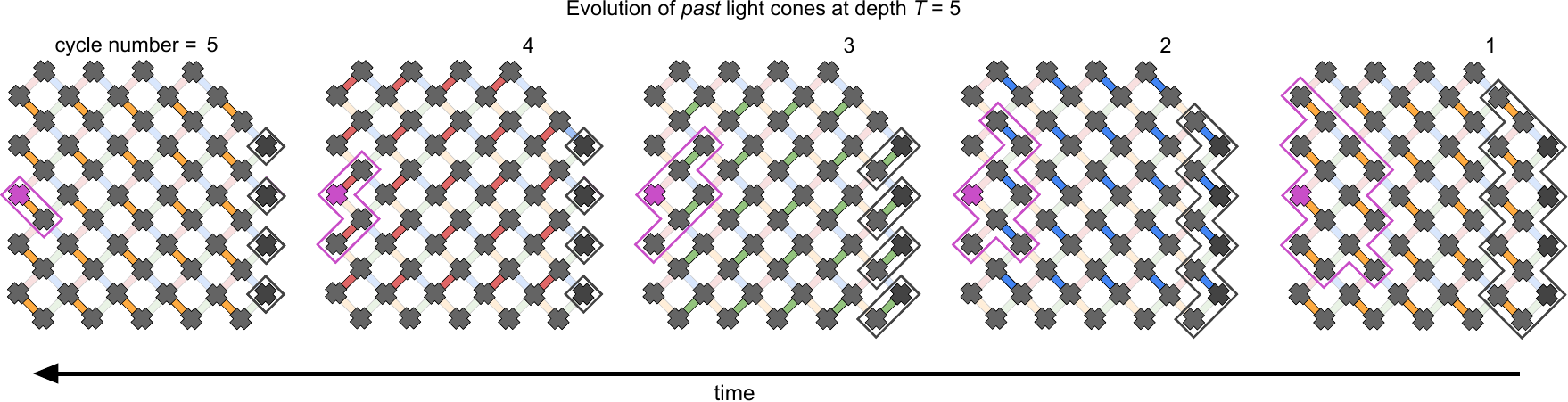} 
\caption{\small{Determination of the past light cones of the probe qubit (pink) and qubits along the right-most column (black) of the $N = 58$ geometry. We begin with the final cycle of two qubit operations at depth $T = 5$ and calculate the past light cone by going backward in time until we reach the first cycle of the circuit. The two subsystems have non-overlapping past light cones and are therefore causally disconnected. \label{fig:lightcones}}}
\end{figure*}

In the process of decoding experimental data as explained above, one obtains classically-simulated values of the probe qubit's Bloch vector $z$-component, $a_m^{z,{\rm sim}}$.
These may be used to compute a simulated order parameter $\zeta^{\rm sim}$ by replacing $a^z_m$ (the ``true'', experimental value) with $a^{z,{\rm sim}}_m$ (the simulated value) in Eq.~\eqref{eq:supp_zeta_def}:
\begin{equation}
    \zeta^{(K),{\rm sim}} = \frac{2}{K} \sum_{i=1}^K |a^{z,{\rm sim}}_{m^{(i)}}| \quad 
    \xrightarrow{K\to\infty}\quad
    \zeta^{\rm sim} = 2\mathbb{E}[|a_m^{z,{\rm sim}}|]
\end{equation}
This simulated order parameter provides information about the ideal (noiseless) dynamics.
In particular it provides evidence of a phase transition in the ideal model, with finite-size crossing behavior, and also provides an estimate of the location of the phase transition.
Data for $S_{\rm proxy}^{\rm sim}$, obtained from the same circuits as Fig.~4 of the main text, is shown in Fig.~\ref{fig:simdecoding}(a). A finite-size crossing at $\rho = \rho_c \simeq 0.72$ is visible. 
The numerically well-established value of the correlation length critical exponent~\cite{Zabalo_exponents_2020}, $\nu \simeq 1.3$, yields a reasonable scaling collapse of the different sizes, Fig.~\ref{fig:simdecoding}(b). This is indicative of a measurement-induced phase transition into the teleporting phase~\cite{Altman2021Teleportation} above $\rho_c$. In the teleporting phase quantum information (aided by classical communication) travels faster than the limits imposed by locality and causality of unitary dynamics. Indeed in the unitary circuit the probe and unmeasured qubits are causally disconnected, with non-overlapping past light cones. This means that there is no qubit in the initial state that may share information with both of them. The formation of the past light cones of the probe (pink) and unmeasured qubits (black) as a function of the cycle number at depth $T = 5$ is illustrated in Fig.~\ref{fig:lightcones}.

\newpage

\section{Approximate decoding}\label{sec:approx_decoding}

Here we describe an approximate decoding algorithm whose computational time scales polynomially in system size $N$, as opposed to the decoding protocol based on exact simulation in Sec.~\ref{sec:decoding}. 
This efficient scaling means the algorithm could be applied on larger quantum processors in the beyond-classical regime.
The algorithm succeeds in the disentangling phase, but fails in the entangling phase. While the protocol's failure is in and of itself a diagnostic of the phase transition, the development of scalable approaches that succeed in both phases and near the critical point in generic models remains an important open problem~\cite{GullansPRL2020, Dehghani2022neural, LiArXiv2022, Barratt_PRLdecoding_2022}. 

\subsection{MPS method}

We adopt the same general approach of Sec.~\ref{sec:decoding}, except the computation of $\tau_m $ is implemented {approximately} via a matrix product state (MPS) method~\cite{Schollwock2011}. MPSs are one-dimensional structures characterized by a cut-off $\chi$, the {\it bond dimension}, that caps the amount of entanglement across each bond: $S \leq \log_2(\chi)$. Classical computational cost scales polynomially in $\chi$, thus exponentially in $S$ (for an exact simulation).
In the disentangling phase, the entanglement within the quasi-1D subsystems used in the simulation is predicted to obey an area-law, $S = O(1)$. Therefore an MPS representation with sufficiently large but finite $\chi$ is expected to provide accurate results even in the $N\to\infty$ limit. Conversely, in the entangling phase, we expect $S$ to grow with decoding radius $r$, $S\sim r$, up to a maximum value $S \sim \sqrt{N}$ (proportional to the linear size of the quasi-1D subsystem). Thus the bond dimension needed for accurate simulation grows super-polynomially with $N$ and accurate simulation becomes unfeasible.

We simulate the 1D monitored circuit of Fig.~\ref{fig:mapping}(b) with an MPS truncated to maximum bond dimension $\chi$ (variable). 
The circuit comprises local unitary gates, projective measurements, and resetting to the $\ket{0}$ state, all operations that can be performed straightforwardly on MPSs. 
The probe qubit is kept at an end of the MPS throughout the simulation.
Note that the spatial range of gates is $\propto T$, thus the method relies on the circuit being shallow (i.e. $T$ being finite). 

\subsection{Results}

\begin{figure}
    \centering
    \includegraphics[width = 0.97\textwidth]{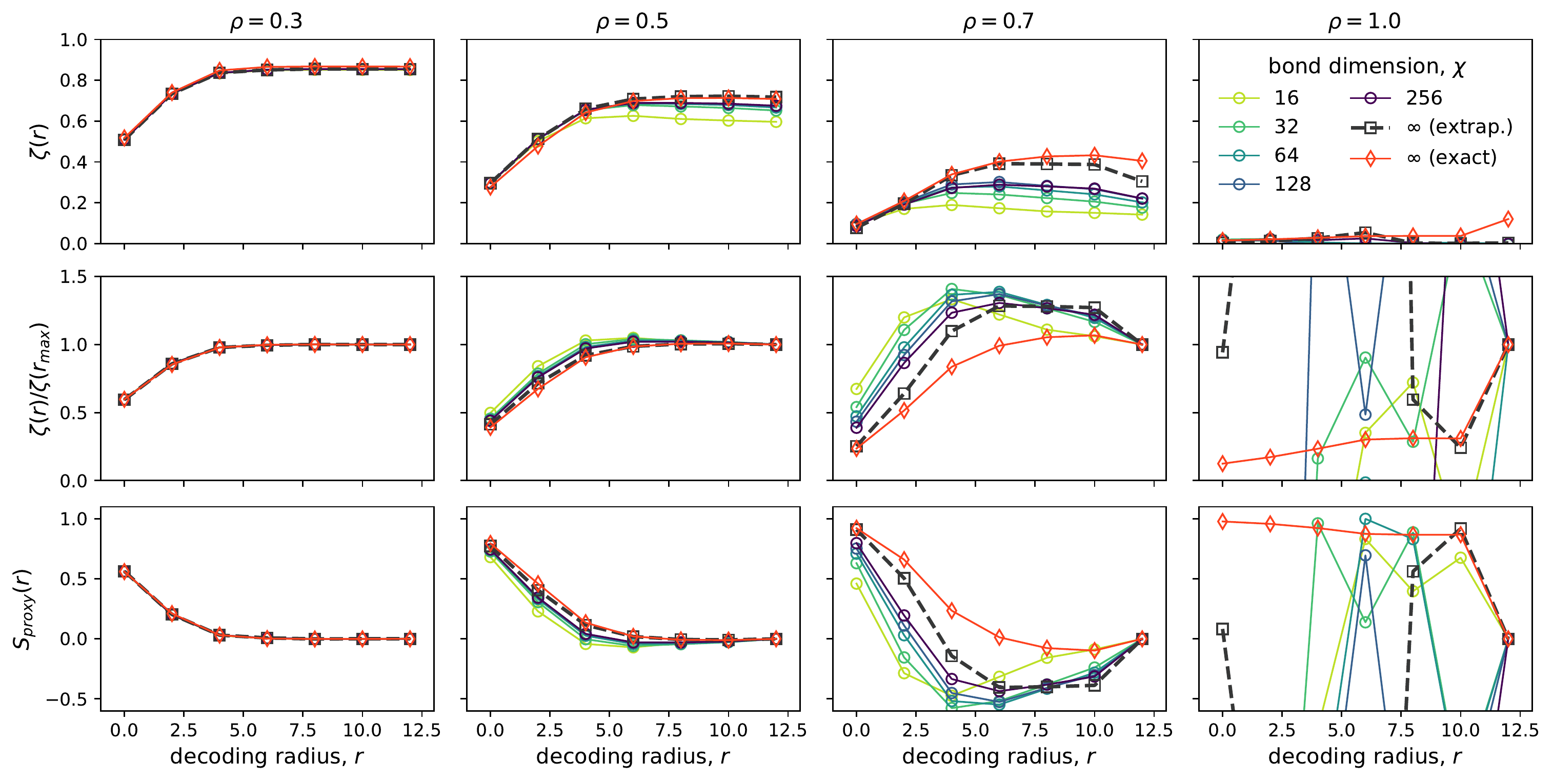}
    \caption{Results of approximate decoding of $N=70$ experimental data based on MPS with bond dimension $\chi = 16, \dots 256$, at four values of $\rho$ across the phase diagram.
    Top row: decoded order parameter $\zeta$ as a function of decoding radius $r$. Middle row: error-mitigated order parameter $\tilde{\zeta}(r) = \zeta(r) / \zeta(r_{\rm max})$. Bottom row: error-mitigated proxy entropy $\tilde{S}_{\rm proxy}(r) = -\log_2[(1+\tilde{\zeta}(r)^2)/2]$. 
    Along with the fixed-$\chi$ data (circles), we show an extrapolation to $\chi \to \infty$ (squares) and a comparison with the results of decoding based on exact simulation (diamonds) as in Sec.~\ref{sec:decoding}. The extrapolation is obtained from fits to an empirical ansatz $\alpha/\log(\chi) + \beta$ on $\chi = 32, \dots 256$. 
    }
    \label{fig:mps_results}
\end{figure}

Results of the approximate decoding algorithm are shown in Fig.~\ref{fig:mps_results} at values of $\rho$ across the phase diagram and for various bond dimensions $\chi$. 
In the disentangling phase, with moderate bond dimension $\chi \leq 256$ the decoding algorithm is fully-converged for $\rho = 0.3$ and nearly-converged for $\rho = 0.5$ (e.g. reasonable extrapolation to $\chi = \infty$ is possible). 
Near the critical point ($\rho = 0.7$) the accuracy of the simulation visibly deteriorates with increasing $r$, as significant truncation errors build up in the simulation. In particular, while $\zeta(r)$ is expected to monotonically increase with $r$, it is found to decrease at large $r$. As a result the mitigated entropy $\tilde{S}_{\rm proxy}(r)$ acquires unphysical negative values.
Finally, deep in the entangling phase ($\rho = 1$) the approximate decoding algorithm fails completely, $\zeta(r)\ll 1$ for all $r$, with no visible improvement at these values of $\chi$. 

\subsection{Effect of classical simulation errors}

In general, the simulated Bloch vector $\vec{a}^{\,{\rm sim}}_m$ will not be perfectly identical to the true (experimental) one $\vec{a}_m$. This is due to unavoidable noise and errors in the experiment, but in the case of approximate decoding may also arise from uncontrolled truncation errors in the classical simulation.
Let us crudely model these effects by assuming that $\tau_m = \textsf{sign}(a^{z, {\rm sim}}_m)$ is correct (i.e. equal to $+\textsf{sign}(a^z_m)$) with probability $q$ and incorrect (i.e. equal to $- \textsf{sign}(a^z_m)$) otherwise, independently of $m$ and circuit realization.
Then Eq.~\eqref{eq:supp_zeta_am} yields
\begin{equation}
    \zeta_{\rm approx}
    = 2\mathbb{E}\left[ a_m^z \textsf{sign}( a^{z,{\rm sim}}_m )\right] 
    = 2(2q-1) \mathbb{E}[|a_m^z|] 
    = (2q-1) \zeta.
    \label{eq:zeta_approx}
\end{equation}
The net result is that the decoded signal is damped by a factor $(2q-1)\leq 1$. 
Further including $r$-dependence in $q$ (as both quantum measurement and classical simulation errors build up with increasing $r$) gives $\zeta_{\rm approx}(r) \sim (2q(r)-1)\zeta(r)$, distorting the signal and causing the error mitigation to fail. This explains the behavior of Fig.~\ref{fig:mps_results} in the entangling phase ($\rho = 1$), where decoding fails. 

\break
\clearpage

\vspace{1em}
\begin{flushleft}
    {\hypertarget{authorlist}{${}^\dagger$}  \small Google Quantum AI and Collaborators}

    \bigskip

    \renewcommand{\author}[2]{#1\textsuperscript{\textrm{\scriptsize #2}}}
    \renewcommand{\affiliation}[2]{\textsuperscript{\textrm{\scriptsize #1} #2} \\}
    \newcommand{\corrauthora}[2]{#1$^{\textrm{\scriptsize #2}, \hyperlink{corra}{*}}$}
    \newcommand{\corrauthorb}[2]{#1$^{\textrm{\scriptsize #2}, \hyperlink{corrb}{\mathsection}}$}
    \newcommand{\xGoogle}{\affiliation{1}{Google Research, Mountain View, CA, USA}}

\newcommand{\xStanford}{\affiliation{2}{Department of Physics, Stanford University, Stanford, CA, USA}}

\newcommand{\xTexas}{\affiliation{3}{Department of Physics, University of Texas at Austin, Austin, TX, USA}}

\newcommand{\xCornell}{\affiliation{4}{Department of Physics, Cornell University, Ithaca, NY, USA}}

\newcommand{\xUMass}{\affiliation{5}{Department of Electrical and Computer Engineering, University of Massachusetts, Amherst, MA, USA}}

\newcommand{\xUCONN}{\affiliation{6}{Department of Physics, University of Connecticut, Storrs, CT}}

\newcommand{\xAU}{\affiliation{7}{Department of Electrical and Computer Engineering, Auburn University, Auburn, AL, USA}}

\newcommand{\xCQC}{\affiliation{8}{QSI, Faculty of Engineering \& Information Technology, University of Technology Sydney, NSW, Australia}}

\newcommand{\xUCR}{\affiliation{9}{Department of Electrical and Computer Engineering, University of California, Riverside, CA, USA}}

\newcommand{\xCU}{\affiliation{10}{Department of Chemistry, Columbia University, New York, NY, USA}}

\newcommand{\xUoCA}{\affiliation{11}{Department of Physics and Astronomy, University of California, Riverside, CA, USA}}

\begin{footnotesize}

\newcommand{\Google}{1}
\newcommand{\Stanford}{2}
\newcommand{\Texas}{3}
\newcommand{\Cornell}{4}
\newcommand{\UMass}{5}
\newcommand{\UCONN}{6}
\newcommand{\AU}{7}
\newcommand{\CQC}{8}
\newcommand{\UCR}{9}
\newcommand{\CU}{10}
\newcommand{\UoCA}{11}

\corrauthora{J. C. Hoke}{\Google,\! \Stanford},
\corrauthora{M. Ippoliti}{\Stanford,\! \Texas},
\author{E. Rosenberg}{\Google,\! \Cornell},
\author{D. Abanin}{\Google},
\author{R. Acharya}{\Google},
\author{T. I. Andersen}{\Google},
\author{M. Ansmann}{\Google},
\author{F. Arute}{\Google},
\author{K. Arya}{\Google},
\author{A. Asfaw}{\Google},
\author{J. Atalaya}{\Google},
\author{J. C.~Bardin}{\Google,\! \UMass},
% ...

\bigskip

\xGoogle
\xStanford
\xTexas
\xCornell
\xUMass
\xUCONN
\xAU
\xCQC
\xUCR
\xCU
\xUoCA



{\hypertarget{corra}{$*$} These authors contributed equally to this work.}



\end{footnotesize}
\end{flushleft}

\bibliography{References.bib}